\documentclass[twocolumn, showpacs, nofootinbib, superscriptaddress, secnumarabic]{revtex4}
\pdfoutput=1                    

\usepackage[utf8]{inputenc}
\usepackage[T1]{fontenc}
\usepackage{mathptmx}
\usepackage{inconsolata}
\usepackage{microtype}

\usepackage{graphicx}
\usepackage{amsmath}
\usepackage{amssymb}
\usepackage[hidelinks]{hyperref}
\usepackage{verbatim}
\usepackage{dsfont}
\usepackage{bm}

\makeatletter
\g@addto@macro\@verbatim\small
\makeatother

\newenvironment{example}{\verbatim}{\endverbatim}
\newenvironment{example*}{\verbatim}{\endverbatim}

\newcommand{\bra}[1]{\left\langle#1\right|}
\newcommand{\ket}[1]{\left|#1\right>}
\newcommand{\braket}[1]{\left\langle#1\right\rangle}

\newcommand{\re}{\text{e}}
\newcommand{\ri}{\text{i}}
\DeclareMathOperator{\imag}{Im}

\newcommand{\mode}{\chi}

\begin{document}

\title{Kwant: a software package for quantum transport}
\date{5 March 2014}
\author{Christoph W.~Groth}
\affiliation{CEA-INAC/UJF Grenoble 1, SPSMS UMR-E 9001, Grenoble 38054, France}
\author{Michael Wimmer}
\affiliation{Instituut-Lorentz, Universiteit Leiden, P.O. Box 9506, 2300 RA Leiden, The Netherlands}
\author{Anton R.~Akhmerov}
\affiliation{Instituut-Lorentz, Universiteit Leiden, P.O. Box 9506, 2300 RA Leiden, The Netherlands}
\affiliation{Department of Physics, Harvard University, Cambridge, Massachusetts 02138 USA}
\author{Xavier Waintal}
\affiliation{CEA-INAC/UJF Grenoble 1, SPSMS UMR-E 9001, Grenoble 38054, France}

\begin{abstract}
  Kwant is a Python package for numerical quantum transport calculations.
  It aims to be an user-friendly, universal, and high-performance toolbox for the simulation of physical systems of any dimensionality and geometry that can be described by a tight-binding model.
  Kwant has been designed such that the natural concepts of the theory of quantum transport (lattices, symmetries, electrodes, or\-bit\-al/spin/electron-hole degrees of freedom) are exposed in a simple and transparent way:
  Defining a new simulation setup is very close to describing the corresponding mathematical model.
  Kwant offers direct support for calculations of transport properties (conductance, noise, scattering matrix), dispersion relations, modes, wave functions, various Green's functions, and out-of-equilibrium local quantities.
  Other computations involving tight-binding Hamiltonians can be implemented easily thanks to its extensible and modular nature.
  Kwant is free software available at \url{http://kwant-project.org/}.
\end{abstract}

\maketitle

\section{Introduction}
Solving the scattering problem is one of the most common and general tasks in condensed matter physics.
Instead of describing states in a closed geometry, one considers the scattering of particles in a finite system coupled (possibly strongly) to infinite leads.
Its solution by itself directly yields the conductance and various other transport properties, but it can also be used as a building block for the calculation of more complicated physical phenomena, such as supercurrent, non-equilibrium density of states at a high voltage bias, or the evaluation of the topological properties of a topological insulator.

The history of numerical simulation of the scattering problem goes back to the early days of mesoscopic physics \cite{lee1981,thouless1981,mackinnon1985} when the first algorithms were developed.
The most popular one of these is the recursive Green's function algorithm (RGF).
Various groups created their own implementations of it, which quickly became an invaluable tool to verify, extend, or even replace the analytical approach even despite being restricted to quasi-one-dimensional geometries and to a particular type of tight-binding Hamiltonian.
Besides quantum transport, the scattering problem naturally emerges in other contexts and many packages with a different  focus (such as density functional theory, or transistor simulations) were developed \cite{transiesta, smeagol, openmx, nanodcal_nanodsim, nemo5, nextnano, nano_tcad_vides}.
Nevertheless, up to now no package existed whose main emphasis is efficiently solving with comparatively little effort the scattering problem for arbitrary single-particle tight-binding Hamiltonians.

\begin{figure}
  \centering
  \includegraphics[width=0.9\linewidth]{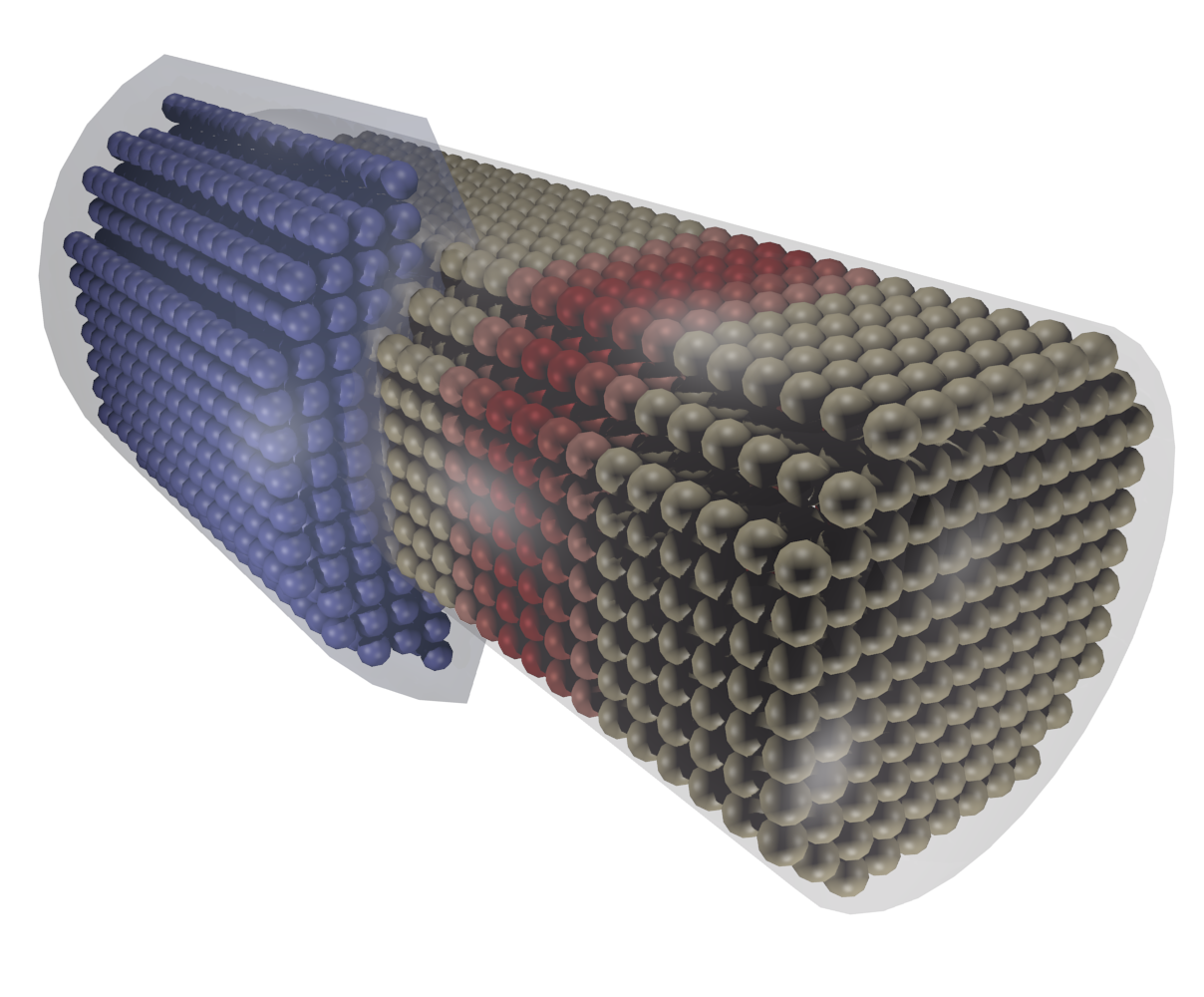}
  \caption{A 3-d model of a semiconducting quantum wire (gray cylinder) simulated with Kwant \cite{mi13}.
    The balls represent lattice sites.
    The red region is a tunnel barrier, used to measure tunneling conductance, the blue region is a superconductor.}
  \label{fig:wire}
\end{figure}

Here we introduce Kwant, a publicly available package that is designed to
\begin{itemize}
\item solve the scattering problem in a robust and highly efficient way,
\item exhibit a high degree of inter-operability with other packages and algorithms from any part of the code, including both defining and solving the scattering problems,
\item support an easy and expressive way to define a broad range of tight-binding systems as required for ex\-plor\-ato\-ry research.
\end{itemize}
Kwant uses highly efficient and robust algorithms that allow one to (i)
significantly outperform the most commonly used recursive Green's function method and (ii) avoid usual instabilities occurring with many commonly used algorithms (for instance in dealing with the evanescent modes of complex electrodes).
Inter-operability removes the need from specialized packages to reimplement the solution of scattering problem, while benefiting from the advanced and efficient algorithms used in Kwant.
Finally, expressiveness is an especially important feature for mesoscopic physics, since it allows one to define a broad range of physical systems using the associated physics concepts directly.
In short, the way one writes down a Hamiltonian in Kwant is very close to what one would write on a blackboard.
The definition of a physical system amounts to writing a simple Python program that operates with physical concepts such as lattices, shapes, symmetries, and potentials.
We hope that the free availability of a user-friendly, generic and high-performance code for quantum transport calculations will help to advance the field by allowing researchers to concentrate more on the physics and to perform computations that were considered out-of-reach due to their complexity.
An example of a device that was simulated with Kwant is shown in Fig.~\ref{fig:wire}: a cylindrical semiconducting wire with spin-orbit interaction, partially covered by a superconductor, used to create Majorana fermions \cite{lutchyn10,oreg10,mourik12}.

\section{Overview of basic concepts of quantum transport}
\label{sec:foundations}

\subsection{Tight-binding systems}
\label{sec:tbsys}
Although Kwant is also suited for finite systems, it mainly targets {\it infinite} systems consisting of a finite scattering region to which a few semi-infinite periodic electrodes are connected.
Within the Landauer-Büttiker formalism these leads act as wave guides leading plane waves into and out of the scattering region and correspond to the contacts of a quantum transport experiment.
The Hamiltonian for such a system takes the form
\begin{equation}
  \mathrm{\hat{\textbf{H}}} = \sum_{i,j} H_{ij} c^\dagger_i c_j,
\end{equation}
where  $c^\dagger_i$  ($c_j$) are the usual fermionic creation (destruction) operators, $i$ and $j$ label the different degrees of freedom of the system, and $H_{ij}$ are the elements of an infinite Hermitian matrix.
Alternatively, the same Hamiltonian can be written in first quantization as
\begin{equation}
  \label{eq:tbham}
  \hat{H} = \sum_{i,j} H_{ij}\ket{i}\bra{j}.
\end{equation}
The degrees of freedom usually take the form $\ket{i} = \ket{\bm{r}\alpha}$, where $\bm{r}$ corresponds to the lattice coordinates of a site, and $\alpha$ labels its internal degrees of freedom.
These can include any combination of spin, atomic orbital, and Nambu electron-hole degree of freedom for superconductivity.
We may express $\hat{H}$ as a sum over all the Hamiltonian fragments
\begin{equation}
  \label{eq:onsite_hop}
  \hat{H}_{\bm{rr'}} = \sum_{\alpha\alpha'}H_{\bm{r}\alpha\bm{r'}\alpha'}\ket{\bm{r}\alpha}\bra{\bm{r'}\alpha'},
\end{equation}
that couple sites $\bm{r}$ and $\bm{r'}$.

Hamiltonians of this form can arise directly from an approximate atomic description of a physical system, in which case sites correspond to atoms or molecules.
Alternatively, a finite-difference discretization of a continuum Hamiltonian also results in a tight-binding Hamiltonian.

\begin{figure}
  \centering
  \includegraphics[width=\linewidth]{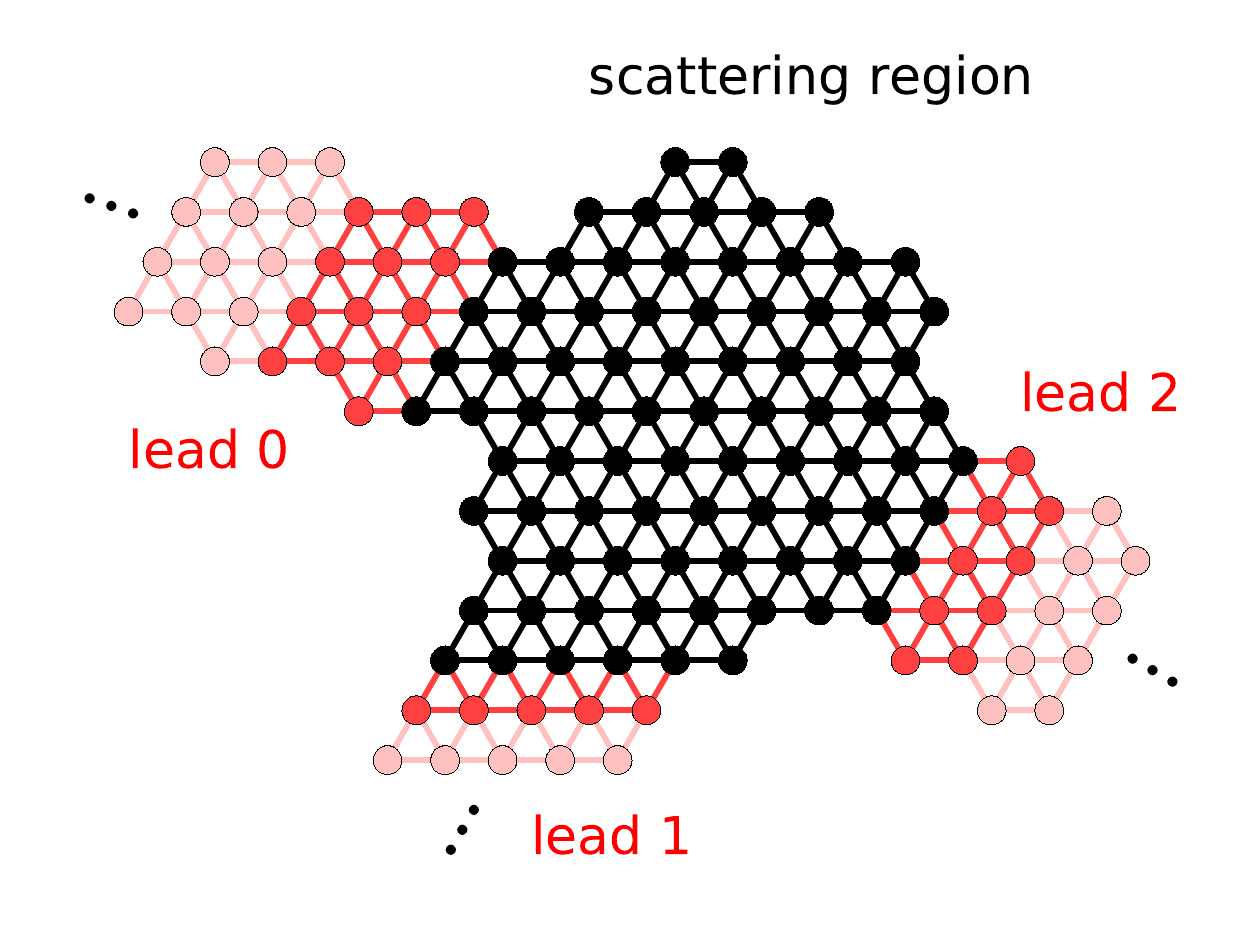}
  \caption{Structure of an exemplary tight-binding system modeled with Kwant.
  Sites belonging to the scattering region are represented by black dots, sites belonging to one of the three semi-infinite leads by red dots.
  Each non-zero off-diagonal Hamiltonian element $H_{\bm{rr'}}$ is shown as a line between site $\bm{r}$ and site $\bm{r'}$.
  Leads consists of an infinite sequence of interconnected identical unit cells.
  In this figure, the unit cells of a lead are drawn in a different shade each.
  Only the first two are shown for each lead.}
  \label{fig:tbsys}
\end{figure}

Kwant represents such Hamiltonians as annotated infinite graphs like the one shown in Fig.~\ref{fig:tbsys}.
Each node of the graph corresponds to a site $\bm{r}$ and is annotated with the typically small Hermitian matrix  $H_{\bm{rr}}$ that is a representation of $\hat{H}_{\bm{rr}}$.
Each edge between sites $\bm{r}$ and $\bm{r'}$ corresponds to a non-zero $H_{\bm{rr'}} = H_{\bm{r'r}}^\dagger$.
The periodicity of the leads allows a finite representation of these infinite objects.

In summary, defining a scattering geometry amounts to defining a graph and the matrices $H_{\bm{rr'}}$ associated with it.


\subsection{Scattering theory}
\label{sec:scattering_theory}
We focus on the wave function formulation of the scattering problem due to its simpler structure compared to non-equilibrium Green's functions, and since Kwant's default solver is based on the wave function approach. The non-equilibrium Green's function formalism is mathematically equivalent to the wave function approach due to the Fisher-Lee relation, however it is less stable \cite{wimmer13}.

Without loss of generality, we can consider the case with a single lead.
This is possible because several leads can always be considered as a single effective lead with disjoint sections.
In the basis in which the sites are ordered according to the reverse distance to the scattering region (scattering region $S$ last, first unit cell of the lead before that, second unit cell before the first one, etc.), the Hamiltonian of such a system has the tridiagonal block form
\begin{equation}
  \label{eq:tbham_inf}
  H =
  \begin{pmatrix}
    \ddots & V_L &  & \\
    V_L^\dagger & H_L & V_L & \\
    & V_L^\dagger & H_L & V_{LS} \\
    & & V_{LS}^\dagger & H_S
  \end{pmatrix},
\end{equation}
where $H_S$ is the (typically large) Hamiltonian matrix of the scattering region $S$.
$H_L$ is the (typically much smaller) Hamiltonian of one unit cell of the lead, while the block submatrix is the Hamiltonian $V_L$ connecting one unit cell of the lead to the next.
Finally, $V_{LS}$ is the hopping from the system to the leads.

We define the wave function of an infinite system as $(\dots, \psi^{\text{L}}(2), \psi^{\text{L}}(1), \psi^S)$,
where $\psi^S$ is the wave function in the scattering region,
and $\psi^{\text{L}}(i)$ the wave function in the $i$-th unit cell away from the scattering region in the lead.
Due to the translational invariance of the leads, the general form of the wave function in them is a superposition of plane waves.
The  eigenstates of the translation operator in the lead take the form
\begin{equation}
  \label{eq:translation_eigenmode}
  \phi_{n}(j) = (\lambda_n)^j \mode_n,
\end{equation}
such that they obey the Schrödinger equation
\begin{equation}\label{eq:lead_equation}
(H_L + V_L \lambda_n^{-1} + V_L^\dagger \lambda_n) \mode_n = E \mode_n,
\end{equation}
with $\mode_n$ the n-th eigenvector, and $\lambda_n$ the n-th eigenvalue.
The normalizability requirement on the wave function reads $|\lambda_n| \leq 1$.
The modes with $|\lambda_n|<1$ are evanescent, the rest are propagating and $\lambda_n=\re^{\ri k_n}$ can be expressed in term of the longitudinal momentum $k_n$ of mode (or channel) $n$.
The propagating modes are normalized according to the expectation value of the particle current, such that
\begin{equation}
\braket{I} \equiv 2\imag\bra{\phi_n(j)} V_L \ket{\phi_n(j-1)} = \pm 1.
\end{equation}
The modes are further sorted into incoming ones $\phi^{\text{in}}_n$ ($\braket{I}=+1$), outgoing ones $\phi^{\text{out}}_n$ ($\braket{I}=-1$) and evanescent ones $\phi^{\text{ev}}_n$ ($\braket{I}=0$).
With these notations, the scattering states in the leads take the form
\begin{equation}
  \label{eq:scat_wave_func}
  \psi_n(i) = \phi^{\text{in}}_n(i) + \sum_{m}S_{mn} \phi^{\text{out}}_m(i)  + \sum_p \tilde{S}_{pn} \phi^{\text{ev}}_p(i),
\end{equation}
and the scattering wave function inside the system
\begin{equation}
  \label{eq:scat_wave_func2}
  \psi_n(0) = \phi^S_n.
\end{equation}
The scattering matrix $S_{nm}$ and the wave function inside the scattering region $\phi^S_n$ are the main raw outputs of Kwant.
Their calculation can be done by matching the wave function in the leads with the one in the scattering region which amounts to inserting the above form of the wave function into the tight-binding equations $H \psi_n = \varepsilon \psi_n$, with $H$ given by Eq.~\eqref{eq:tbham_inf}.

Examples of transport properties that can be obtained from the scattering matrix include conductance, shot noise, spin currents, Peltier and Seebeck coefficients and many other quantities.
For instance, the differential conductance $G_{ab}=dI_a/dV_b$ (where $a$ and $b$ label two electrodes) is given by the Landauer formula
\begin{equation}
  \label{eq:Landauer}
  G_{ab}=\frac{e^2}{h} \sum_{n\in a, m\in b} |S_{nm}|^2.
\end{equation}
The internal properties of the system such as local density of states or  current density can be obtained from the $\phi^S_n$ using the general relation
\begin{equation}
  \label{eq:Lesser}
 \braket{c^\dagger_i c_j} =\int \frac{dE}{2\pi} \sum_n f_n(E)  \phi^S_n(j)  \left[\phi^S_n(i)\right]^*,
 \end{equation}
where $f_n(E)=1/[1+e^{(E-\mu_n)/kT_n}]$ is the Fermi function of the lead to which channel $n$ is associated.

\section{The design of Kwant}
\label{sec:design}
As explained in the introduction, we have designed Kwant for performance and interoperability on the one hand, and flexibility and ease of use on the other.
Combining all of these requirements is a nontrivial task since flexibility and ease of use can be best achieved using features of a high-level language (Python in our case), while performance and interoperability require a universal low-level language interface as well as simple data structures.

This apparent contradiction is resolved once we notice that while the time necessary for defining the tight-binding Hamiltonian will scale linearly with system size (if implemented efficiently), solving the scattering problem or applying any other relatively complicated numerical algorithm will scale less well.
This allows us to separate the work into two phases:
During the first the tight-binding Hamiltonian is constructed in Python using concepts that are natural in physics such as lattices, shapes, and functions of coordinates (e.g.\ an electrostatic potential $V = \tanh(\alpha x)$ or a hopping in presence of magnetic field, $t = t_0 \exp[i B_z (x_1 - x_2) (y_1 + y_2)]$).
Subsequently, the Hamiltonian is prepared for the second phase by transforming it into a low-level representation.
During the following second phase, this optimized representation is used as input for high-performance numerical calculations.

As we show later in Sec.~\ref{sec:benchmark}, this two-phase approach indeed does not cause any significant drop in performance.
On the contrary, using the \emph{nested dissection} algorithm \cite{George73} implemented in sparse linear algebra libraries, such as MUMPS \cite{mumps1, mumps2}, allows Kwant to significantly outperform a reference implementation of the RGF algorithm written in pure C.
An additional advantage of the separation of defining tight-binding systems and solving the scattering problem is that the default implementations of either of the two phases can be substituted by other ones that are more adapted to a specific problem.

\subsection{Defining tight-binding systems}
\label{sec:defin-tight-bind}

The most natural way to think about a finite tight-binding system is to consider it as a mapping from the vertices and edges of a graph to the corresponding values of the Hamiltonian for the sites and hoppings.
In Kwant such a mapping is represented by a \texttt{Builder} object.
Its implementation as a hash table (Python dictionary) allows to efficiently add or remove sites and hoppings that are present in the system, thereby changing the geometry of the system incrementally.

\begin{figure}
\includegraphics[width=\linewidth]{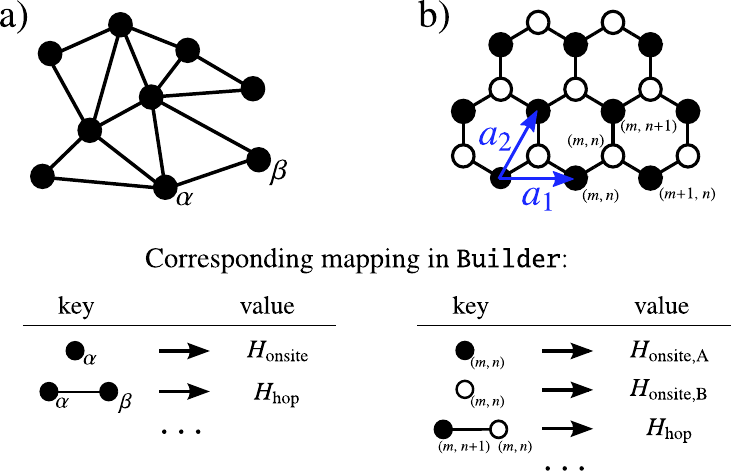}
\caption{Examples of tight-binding systems and a pictorial representation of the corresponding mapping in Kwant.
  (a) Example of an irregular tight-binding lattice with a single site family (filled circles) and site tags $\alpha, \beta, \dots$.
  (b) Example of a regular honeycomb lattice.
  The two sublattices A and B are mapped to two site families (filled and open circles), and site tags are given as tuples of integers (lattice indices).
}\label{fig:design}
\end{figure}

Sites can be often classified by type of atom or the lattice to which they belong.
We represent this in Kwant by defining a site to be a combination of a \emph{family} and a \emph{tag}.
The site family identifies the class of the site, while the site tag identifies a unique site within this family.
An example of a tight-binding system with only a single site family is shown in Fig.~\ref{fig:design}(a).
The \texttt{Builder} object then maps every site (identified by tag $\alpha$) to a \emph{value}, in this case the onsite Hamiltonian term.
Hoppings are represented as pairs of sites and mapped to the corresponding hopping Hamiltonian.
Note that in this example, the site tag could be any object, such as an integer or a string.

This way of defining a tight-binding system is completely universal.
However, in the common case when many sites belong to the same crystal structure the notion of site family becomes more useful.
In that case we can define a site family to represent each of the sublattices and the site tags to be tuples of integer coefficients $(n_1, \dots, n_{d})$ that describe the site position coefficients in the basis of the Bravais lattice vectors $\bm{a}_1, \dots, \bm{a}_{d}$, with $d$ the dimensionality of the lattice.
This enables us to also use operations in real space, since the site tags are directly related to the real-space position $\bm{x} = \sum_i n_i \bm{a}_i$.
For example we may operate with all sites of a lattice that lie within a certain geometrical shape, or make the Hamiltonian values depend on the real space position of the sites.
This approach applied to a honeycomb lattice with two site families corresponding to its two sublattices is shown in Fig.~\ref{fig:design}(b).

A tight-binding model on a regular lattice typically contains only a few `kinds' of hoppings, with hoppings of a single kind being related to each other by lattice translations.
In order to make use of that, Kwant allows to manipulate all hoppings of a single kind in a single operation.

While it would be sufficiently general to only allow complex constants as elements of the Hamiltonian, Kwant allows two important generalizations motivated by common usage.
First, it is natural to identify several degrees of freedom that occupy the same position in space, such as orbitals, with a single site.
In this case the values of the onsite Hamiltonians and the hoppings become matrices, as defined in Eq.~\eqref{eq:onsite_hop}.
Second, the Hamiltonian matrix elements can be often seen as functions of position and other parameters on which the calculation may depend.
To accommodate this, Kwant allows the values of a Hamiltonian to be given by program subroutines that are evaluated at a later stage.

The final feature of \texttt{Builder} that we are going to discuss is symmetry support.
In order to reduce the amount of book-keeping, Kwant enforces the Hermiticity of tight-binding systems: $H_{\bm{rr}'} = H^\dagger_{\bm{r}'\bm{r}}$.
In addition, tight-binding systems in Kwant are allowed to have a certain real space symmetry that is automatically enforced.
Builders with a translational symmetry may be attached to other builders, thereby forming the leads in a scattering geometry, as described in Sec.~\ref{sec:scattering_theory}.

The details of the \texttt{Builder} interface are described in Appendix \ref{app:builder}.

\subsection{Low-level representation of tight-binding systems}
\label{sec:low-level-repr}

The mapping format used by \texttt{Builder} objects is very useful for defining tight-binding systems, but it is a rather poor choice for performing computations.
First of all, much of it is specific to Python, and hence hard to interface with code written in other languages.
It is also not memory-efficient, and does not allow to easily calculate certain crucial properties of the tight-binding system, such as the Hilbert space dimension.
In order to avoid these problems, we define a low-level representation of a tight-binding system suitable for efficient calculations.
In essence this representation is a sparse graph of numbered sites and an array of values of sites and hoppings of this graph.
The aspects that distinguish such a system from a regular sparse matrix are that the values may be functions, and that semi-infinite leads may be attached to it.

A low-level system may be created from a \texttt{Builder} by \emph{finalizing} it.
In the case where Kwant is interfaced with an external package, which performs e.g.\ a DFT calculation, a low-level system may also be defined directly by the other package, fully avoiding the usage of Kwant or even Python for defining the system.
Once a low-level system has been created, it can be used to perform numerical calculations.
Kwant uses low-level systems as an input to quantum transport solvers that calculate various quantities of interest for tight-binding systems with multiple attached semi-infinite leads.
Kwant contains efficient implementations of algorithms for computing the scattering matrix, the retarded Green's function, the scattering wave functions, modes in the leads, and the local density of states.

Since low-level system is an universal format, any computation with tight-binding Hamiltonians can be trivially adapted to use Kwant low-level systems as input.
If, for example, one is interested in the eigenstates of a quantum system, the full Hamiltonian matrix can be requested from the low-level system for a specific set of parameters and passed on to a standard eigenstate calculation routine as provided for example by ARPACK \cite{lehoucq97} (bundled for Python by SciPy \cite{jones01}).
On the other hand, the fact that the Hamiltonian values are implemented as functions allows to naturally integrate Kwant with a Poisson solver, and implement Hatree-Fock mean field calculations.

The solving phase and low-level systems are described in more detail in Appendix~\ref{app:solvers}.

\section{Comparison with other quantum transport packages}
The quantum transport problem appears in many physical settings, and hence it is not surprising that it is addressed by various software packages from different domains.
In particular, there are quite a few packages, including commercial ones, for computing transport in molecular junctions.
Examples of these packages are TranSiesta/Atomistix Toolkit \cite{transiesta}, SMEAGOL \cite{smeagol}, OpenMX \cite{openmx} or nanodcal/nanodsim \cite{nanodcal_nanodsim}.
These combine density functional theory (DFT) with the non-equilibrium Green's function technique.
Another group of codes that deal with the transport problem is mainly geared towards the simulation of transistors on the nano-scale.
It includes NEMO5 \cite{nemo5}, nextnano \cite{nextnano}, or NanoTCAD Vides \cite{nano_tcad_vides}, and TB\_Sim \cite{tb_sim}.
These packages focus on more complicated physical effects, and often go beyond the scope of Kwant by including phonon effects, Boltzmann transport, or self-consistent electrostatic potential calculation.
However, all these packages are specialized to a certain class of tight-binding systems, and extending them to include an additional physical effect, such as superconductivity is often impossible or requires a lot of work.

In contrast, Kwant focuses on generality in order to allow simulating the broad variety of complex models and geometries that are encountered in mesoscopic physics.
There exist several private codes for solving the scattering problem (for example Refs.~\cite{Wimmer2009, Schomerus_code, SanJose_code}) in addition to the publicly available KNIT package \cite{Kazymyrenko2008} in which one of us was involved.
To the best of our knowledge Kwant significantly outperforms these packages since it uses the nested dissection algorithm instead of RGF (see Sec.~\ref{sec:benchmark} for details).
Another advantage of Kwant, as described in the Sec.~\ref{sec:design}, is that it provides both an advanced set of tools for defining tight-binding models, and a universal interface that allows to interact with other codes.

\section{Illustration of Kwant usage: universal conductance fluctuations in a
  quantum billiard}
\label{sec:billiard}
In order to illustrate how various aspects of Kwant work together when applied to a scattering problem we turn to the classic example of deterministic chaos in a stadium billiard.
Despite their regular shape, stadium billiards are not integrable, showing an irregular density of states inside the scattering region, and universal conductance fluctuations.
Thanks to Kwant's expressiveness, the Python program that defines the billiard system, performs numerical calculations, and creates two figures is less than 30 lines long.
In the following the complete program is presented together with explanations.

\begin{figure}
  \centering
  \includegraphics[width=\linewidth]{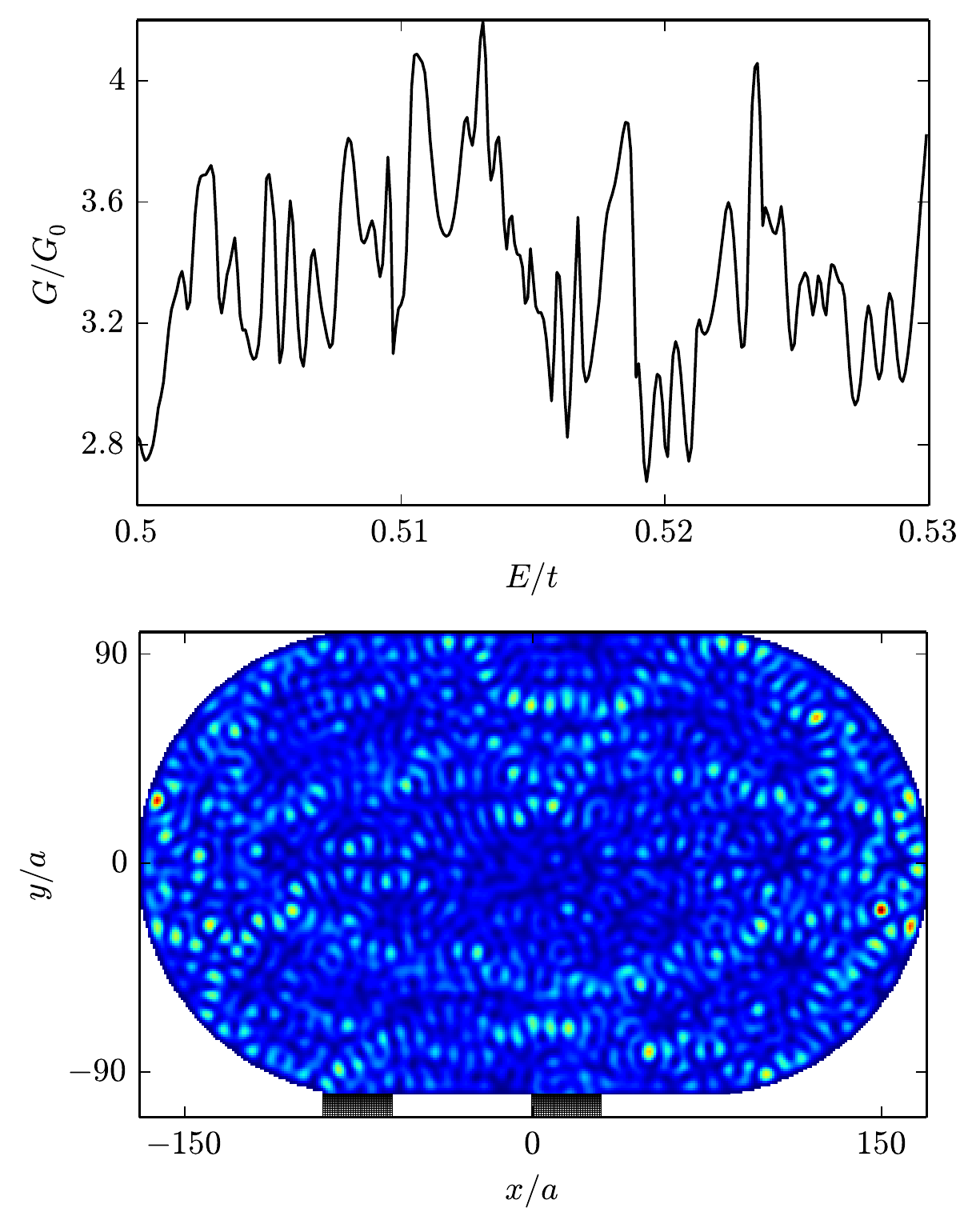}
  \caption{Top panel: dependence of conductance of a stadium billiard on energy showing universal conductance fluctuations.
    Bottom panel: Color plot of the local density of scattering states in the same billiard at a given energy.
    The first ten unit cells of the attached semi-infinite lead are visible as a shaded rectangle below the billiard.
    This figure has been generated by the Kwant script shown in Sec.~\ref{sec:billiard}.
  }
  \label{fig:billiard}
\end{figure}

The first step is to make Kwant's functionality available within Python,
\begin{example}
import kwant
\end{example}
As described in Sec.~\ref{sec:defin-tight-bind}, we need to create the builder object that will contain the information about the system being constructed.
We also create the lattice that is used.
\begin{example}
sys = kwant.Builder()
sqlat = kwant.lattice.square()
\end{example}
We continue with defining a function that specifies the shape of the the scattering region.
Given a point $(x, y)$ this function returns \texttt{True} if the point is inside the shape and \texttt{False} otherwise.
\begin{example}
def stadium(position):
    x, y = position
    x = max(abs(x) - 70, 0)
    return x**2 + y**2 < 100**2
\end{example}
We proceed to add these sites to the scattering region and set the corresponding values of the onsite potential to $-4t$ (we use $t=-1$).
\begin{example}
sys[sqlat.shape(stadium, (0, 0))] = 4
\end{example}
The expression \texttt{sqlat.shape(stadium, (0, 0))} represents all the sites of the lattice that belong to the stadium (provided they can be reached from the central point $(0, 0)$).

We then set the hopping Hamiltonian matrix elements between nearest neighbors
to $t=-1$:
\begin{example}
sys[sqlat.neighbors()] = -1
\end{example}
The scattering region is now fully defined.

To complete the scattering problem, as explained in Sec.~\ref{sec:scattering_theory}, we need to define the leads.
The procedure for a lead is very similar to that for the scattering region.
The only difference between the two is that upon creation of each lead its symmetry is specified.
All further operations with the lead will automatically respect this symmetry: For example, adding a single site to the lead will also add all the image sites under the symmetry as well.
Thus, in order to provide enough information, specifying the structure of a single unit cell is sufficient.
We construct two leads by defining the sites which belong to a unit cell of each lead, and attach them to the scattering region.
\begin{example}
lead_symmetry = kwant.TranslationalSymmetry([0, -1])
for start, end in [(-90, -60), (0, 30)]:
    lead = kwant.Builder(lead_symmetry)
    lead[(sqlat(x, 0) for x in range(start, end))] = 4
    lead[sqlat.neighbors()] = -1
    sys.attach_lead(lead)
\end{example}
This finishes the definition of the scattering problem.

The next required step, as explained in Sec.~\ref{sec:low-level-repr}, is to transform the system into a form suitable for efficient numerical calculations:
\begin{example}
sys = sys.finalized()
\end{example}
%
We can now use the Kwant solvers to obtain various physical observables.
We compute the conductance, given by Eq.~\eqref{eq:Landauer}, and the local density of states using Eq.~\eqref{eq:Lesser}:
\begin{example}
energies = [0.5 + 1e-4 * i for i in range(300)]
conductances = [kwant.smatrix(sys, en).transmission(1, 0)
                for en in energies]

local_dos = kwant.ldos(sys, energy=.2)
\end{example}
The function \texttt{kwant.smatrix} return the scattering matrix of the system at a given energy.
This scattering matrix is then used to calculate the transmission from one lead to another.
The function \texttt{kwant.ldos} returns the local density of states in the scattering region.

We finish the program by plotting the calculated data as shown in Fig.~\ref{fig:billiard}:
\begin{example}
from matplotlib import pyplot
pyplot.plot(energies, conductances)
pyplot.show()
kwant.plotter.map(sys, local_dos, num_lead_cells=10)
\end{example}

Here, in order to make the leads visible, we have plotted the first ten lead unit cells.

The resulting plots (see Fig.~\ref{fig:billiard}) show the universal conductance fluctuations of the stadium billiard and its irregular, chaotic density of states.
We see that defining the scattering problem and calculating the relevant physical observables is transparent, logical, and involves a minimal number of steps.

\section{A full-scale application: Hanle effect in a graphene-based non-local spin valve}
\label{sec:valve}
\begin{figure}
  \centering
  \includegraphics[width=\linewidth]{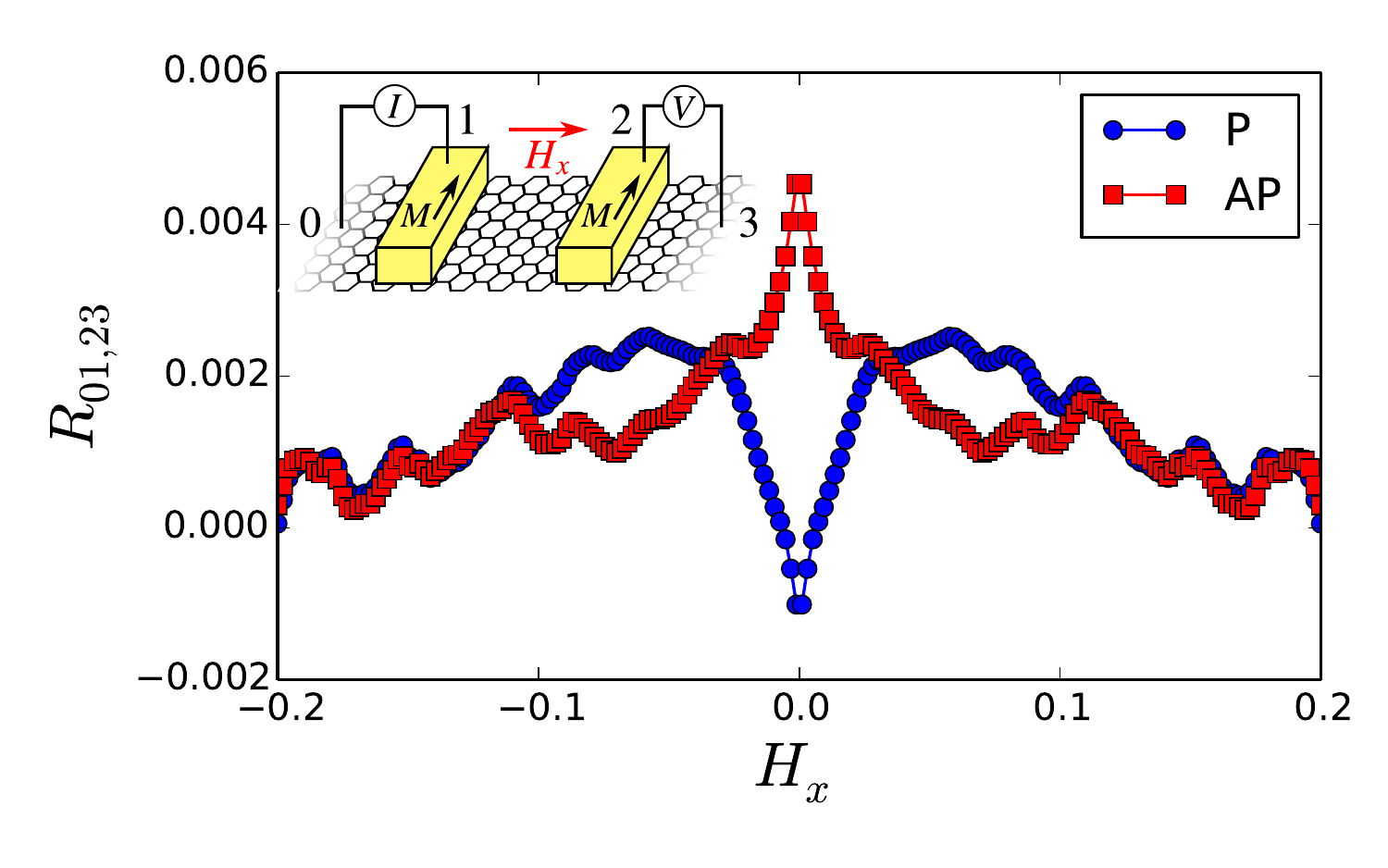}
  \caption{Hanle effect in a graphene-based spin valve.
Blue circles (red squares)  correspond to the non-local resistance $R_{01,23}=V/I$ as a function of a perpendicular magnetic field $H_x$ for the parallel (anti-parallel) configuration.The model parameters are $\beta=0.5$, $t=0.2$ and $W=0.4$.
The graphene ribbon contains around $7000$ carbon atoms.
Inset: schematic of the four-terminal non-local spin valve including the two top magnetic contact.}
  \label{fig:valve}
\end{figure}

We continue with a simulation of a graphene-based non-local spin valve as sketched in the inset of
Fig.~\ref{fig:valve}.
The device consists of a graphene nanoribon where the two sides serve as contacts $0$ and $3$.
Two additional magnetic metallic leads ($1,2$) are deposited on top of the nanoribon.
This setup allows to measure the non-local resistance $R_{01,23}=V/I$ of the device: a current $I$ is passed through contacts $0$ and $1$ and the voltage $V$ is measured between $2$ and $3$.
Such a device allows to measure a ``pure'' spin signal since no electrical current is flowing through the electrodes $2$ and $3$ where the voltage
drop is measured, and was studied experimentally recently \cite{Tombros07,Roche14}.

We model the system using a tight-binding Hamiltonian,
\begin{equation}
  \mathrm{\hat{\textbf{H}}} = \sum_{\langle ij\rangle,a,ss'} t_a^{ss'} \ket{i,s}\bra{j, s'}
 +\sum_{i\in G,ss'} V_i^{ss'} \ket{i,s}\bra{i, s'},
\end{equation}
where the sum over $\langle ij\rangle$ is restricted over nearest neighbours and the corresponding hopping amplitude $t_a^{ss'}$ takes different values inside the graphene layer ($a=G$), the magnetic electrodes ($a=F$) and at the graphene-ferromagnet interface ($a=GF$). $s$ and $s'$ denote spin indices. The metallic magnetic leads
are modeled using a cubic lattice; it is attached to the honeycomb lattice
of graphene by adding a hopping $t_{GF}$ from each lattice point in the last
slice of the cubic lattice to the nearest atom in the graphene lattice.

We take $t_F=t_G=1$ while the spin-filtering due to the presence of the magnetic electrodes is included in the interfacial hopping,
\begin{equation}
t_{GF}= (t/2) ( 1 +\beta \sigma_z ),
\end{equation}
where $-1 < \beta < 1$ characterizes the spin polarization of the graphene-ferromagnet interface.
Lastly, the graphene onsite potential contains static disorder plus an in-plane magnetic field $H_x$ perpendicular to the magnetization of the electrodes
\begin{equation}
V_i = W v_i + H_x \sigma_x
\end{equation}
where the $v_i$ are random numbers uniformly distributed inside $[-0.5,0.5]$, $W$ characterizes the strength of the disorder potential and $\sigma_x,\sigma_z$ are the Pauli matrices.

This is the minimal model that allows to simulate the Hanle effect in
a lateral spin valve.  Additional ingredients, such as magnetic
disorder, could be added to introduce a finite spin-diffusion length
in the system.


Even though we present a minimal model, it is far more complex than the
two-terminal geometries typically studied in numerical simulations.
In particular, it is necessary to study a four-terminal geometry involving
different Bravais lattices, and to include the spin degree of freedom.
Kwant's design allows a natural implementation of this system: The different
Bravais lattices are represented by different site families, and the
spin degree of freedom by matrix Hamiltonians associated with sites and hoppings in
\texttt{Builder}. In addition, the solver for computing the scattering
matrix is general enough to allow for arbitrary geometries with an arbitrary
number of contacts.


Apart from native Kwant features, the present example also benefits
from the fact that Kwant is a Python package and as such can be easily embedded and extended
by custom code. For example, the functionality for connecting the cubic lattice leads
with their graphene substrate is not included in Kwant and must be custom-coded by the user.
This is easily achieved by combining coordinate information provided by Kwant
with a $kd$-tree (a data structure designed to find the nearest lattice point)
provided by the SciPy package \cite{jones01}. Also, the calculation of the
non-local resistance requires solving a $3\times 3$ linear equation,
which is done in one line of code using a call to the NumPy package
\cite{oliphant07}. Finally, a few additional lines of code allow
to parallelize the script for a multi-core workstation or cluster.


Fig.~\ref{fig:valve} shows the result of a numerical simulation using Kwant:
the non-local resistance as a function of $H_x$ for the two configurations where the magnetizations are parallel $P$ and anti-parallel $AP$ (the latter is obtained by setting $\beta \rightarrow -\beta$ in one of the magnetic electrodes).
We find a typical Hanle signal: when the magnetic field is increased, the spin precesses around the x axis resulting in a change of the sign of the spin-dependent signal $\Delta R =R_P -R_{AP}$.
Since different trajectories have different lengths, the precession angle is spread over a finite window.
Eventually, this makes $\Delta R$ vanish at large magnetic fields.

Hanle precession is typically described within a semi-classical diffusion description, and the current study is to our knowledge the first to take into
account quantum coherence. In fact, the full quantum model presented here
allows to go beyond a semi-classical description and study the effect of
phase coherence and/or ballistic propagation ($W=0$).

\section{Benchmark}
\label{sec:benchmark}
We now show a comparison of the performance of Kwant with a C implementation of the RGF algorithm applied to a prototypical quantum transport problem: the calculation of the conductance of a square tight-binding system.
The system consists of $L \times L$ single-orbital sites that belong to a square lattice.
On two opposite sides leads of width $L$ are attached.
The energy was kept fixed, so that the number of propagating modes in the leads was proportional to $L$.
The measurements were performed on a computer with an Intel i5-2520M 64-bit processor and 8 GiB of main memory running a variant of GNU/Linux with single-threaded OpenBLAS \cite{zhang2012}.

\begin{figure}
  \centering
  \includegraphics[width=\linewidth]{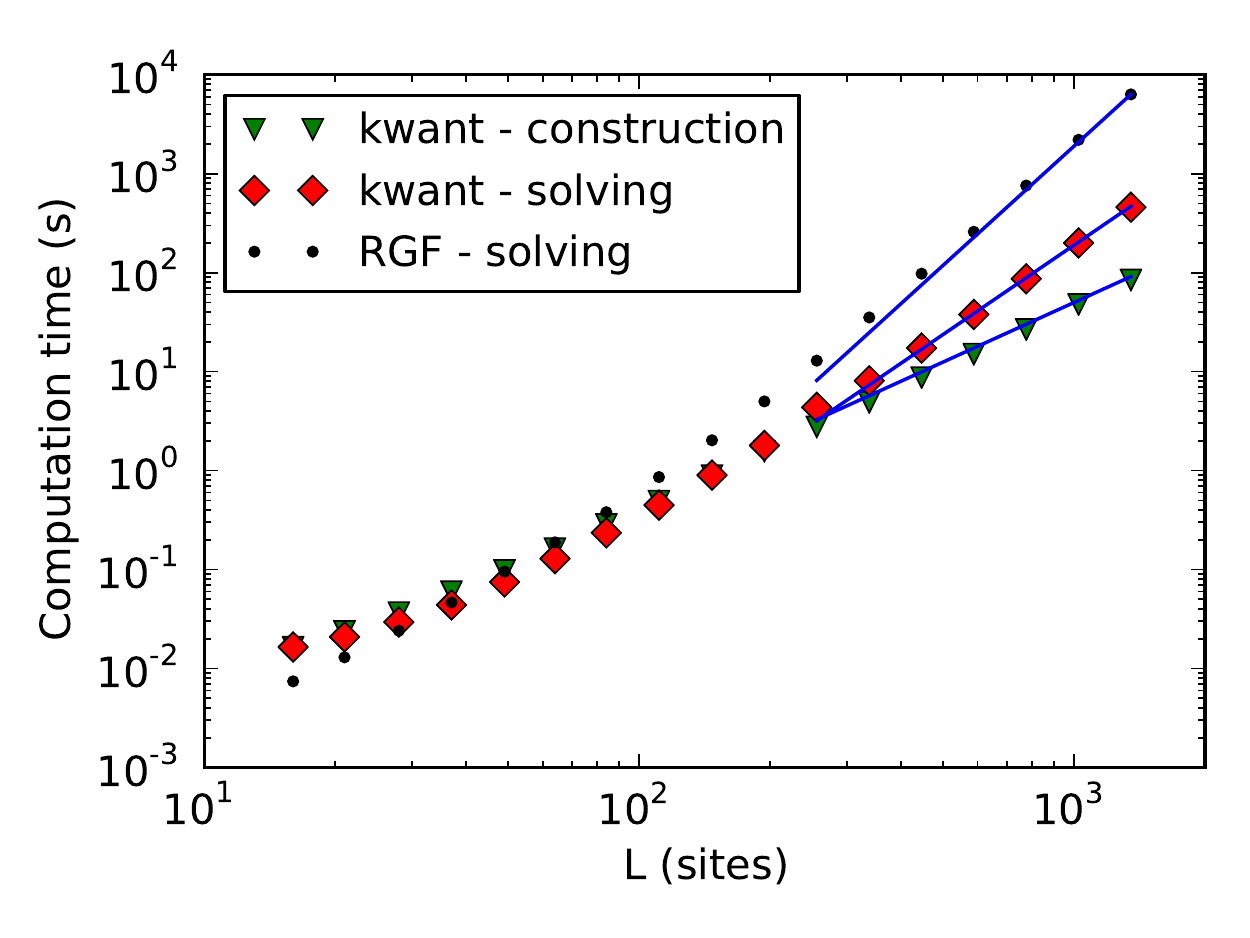}
  \caption{Time used for calculating the conductance of a square tight-binding system of side length $L$.
    Triangles: construction (including finalization) of the system with Kwant.
    Diamonds: solving with Kwant's MUMPS-based solver.
    Dots: solving with an efficient C implementation of the RGF algorithm.
    The lines show the theoretically expected scaling for large $L$: $O(L^2)$ for construction, $O(L^3)$ for solving with the MUMPS-based solver, and $O(L^4)$ for solving with RGF.}
  \label{fig:time}
\end{figure}

Figure~\ref{fig:time} shows the dependence of the running time on $L$ and compares Kwant with an alternative code written entirely in C (using the same BLAS and LAPACK) that implements the RGF method.
One can see that for large system sizes, Kwant with the MUMPS-based solver is up to ten times faster than the C RGF code.
For small systems the situation is inverted but less dramatic, the cross-over occurs around $L = 50$.
It is evident that for small systems Kwant's construction step takes up a considerable fraction of the total time.
The much simpler RGF code only supports quasi-1-d geometries and therefore requires no system construction in the sense of Kwant.
Construction is typically performed considerably less often (one time per geometry) than solving (once per geometry and set of parameters).

\begin{figure}
  \centering
  \includegraphics[width=\linewidth]{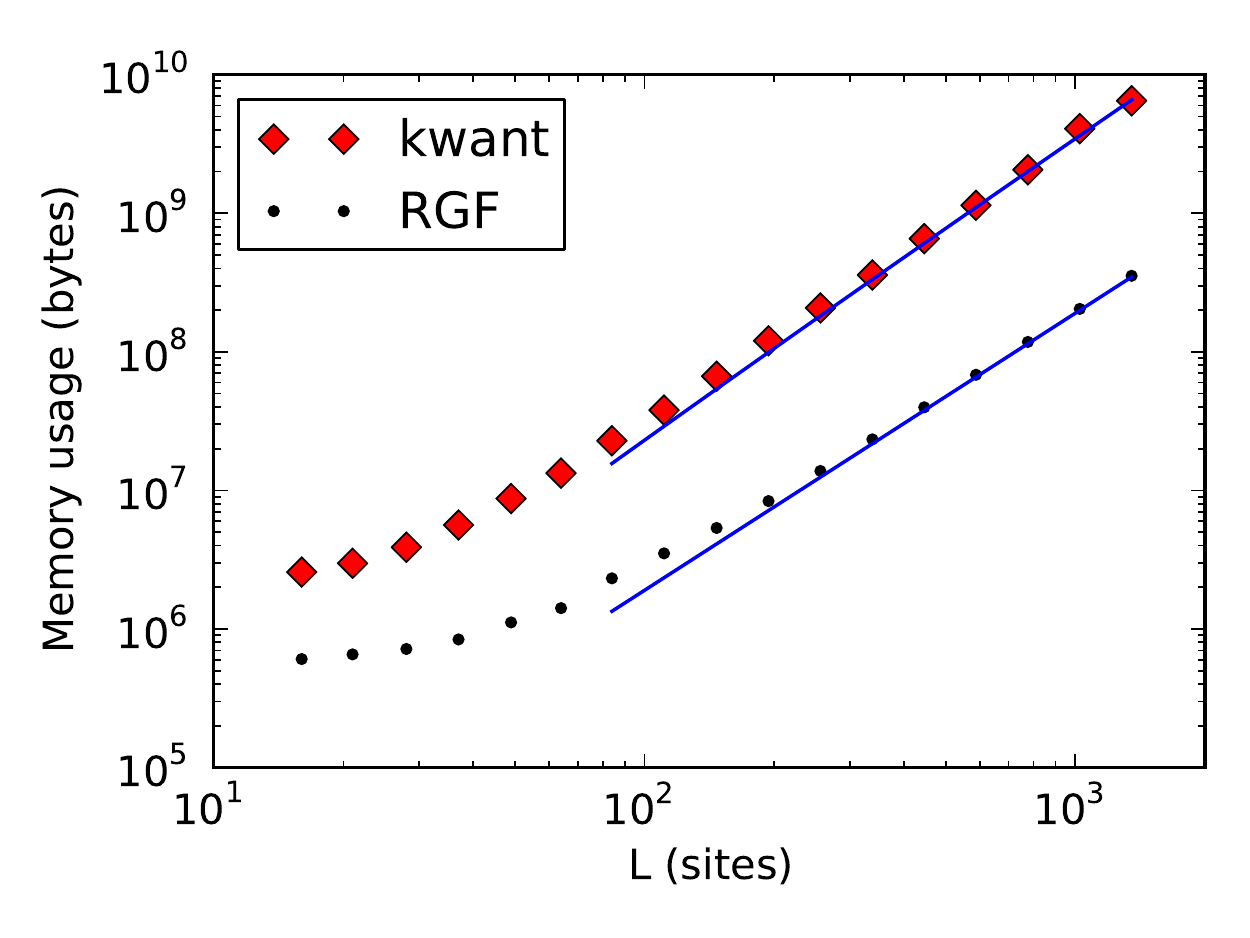}
  \caption{Memory used for calculating the conductance of a square tight-binding system of side length $L$.
    Diamonds: Kwant with the MUMPS-based solver.
    Dots: an efficient C implementation of the RGF algorithm.
    The continuous curves show the theoretically expected scaling for large $L$: $O(L^2 \log L)$ for MUMPS-based Kwant solver, and $O(L^2)$ for RGF.}
  \label{fig:memory}
\end{figure}

Figure~\ref{fig:memory} compares the memory footprint of Kwant's MUMPS-based solver with that of an RGF solver implemented in C.
Evidently, increased memory usage is the price for the superior speed of Kwant's solver.
Note, however, that with main memory sizes commonly available today even on low-end computers (a few GiB) the MUMPS-based solver is able to tackle systems of more than $10^6$ sites.
To be able to handle even larger systems RGF has been implemented as well.
(It is not yet part of the public release of Kwant as of version 1.0.)
The shown memory consumption is the additional memory required for the computation on top of a basic constant requirement (71 MiB for Kwant, 35 MiB for the RGF solver) that is small by today's standards but whose inclusion would have obstructed the power-law character of the curves.
It was measured as the increase of the maximum ``resident set size'' taken up by a given computation over that reported for an empty run of the respective software.

\section{Conclusion}
\label{sec:conclusion}
The Kwant project has a double objective.
First, to gather high-performance algorithms that are useful in the field of quantum transport.
Second, to provide a simple and clear but powerful user interface for defining and working with tight-binding models.
We refrained from designing such an interface from scratch (an approach that usually does not age well) but rather chose to extend an existing language (Python) with new capabilities.
In this approach, the usual input files present in most scientific software disappear and are replaced by small programs.
This results in more flexibility, since tight integration with other packages becomes possible, as well as pre- and post-processing of the data.
We believe that such a modern approach to scientific programming has a strong impact on the usefulness of a code.

This paper describes version 1.0 of Kwant, the first version released to the public.
Kwant is a work in progress and there are many ideas for improvements collected in a to-do-list that is available with the source code.
For instance, more solvers could be added and Kwant's low-level system format could be modified to allow general symmetries.

Kwant is free (open source) software \cite{freesoftware} (distributed under the liberal ``simplified BSD license''), and we hope that the project's website \url{http://kwant-project.org/} and mailing list \url{http://kwant-project.org/community} will develop into a hub for a community of users and developers.
Contributions to Kwant as well as the sharing of related code modules are welcome.

\section*{Acknowledgements}
We gratefully acknowledge contributions by D.~Jaschke, and J.~Weston, as well as discussions with P.~Carmier, M.~Diez, C.~Fulga, B.~Gaury, B.~van~Heck, D.~Nurgaliev, and D.~Wecker.
We thank C.~Gohlke for the creation of Windows installers.
ARA and MW were supported by the Dutch Science Foundation NWO/FOM and by the ERC Advanced Investigator Grant of C.~W.~J.~Beenakker who enthusiastically supported this project.
ARA was partially supported by a Lawrence Golub fellowship.
CWG and XW were supported by the ERC Consolidator Grant MesoQMC as well as by the E.U. flagship graphene.
XW also acknowledges support from the STREP ConceptGraphene.

\appendix

\renewcommand{\thesection}{\Alph{section}}
\renewcommand{\thesubsection}{\thesection.\arabic{subsection}}
\renewcommand{\thesubsubsection}{\thesubsection.\arabic{subsubsection}}

\section{Programming interface of builder objects}
\label{app:builder}

\subsection{Sites and site families}
Site objects are Kwant's abstraction for the sites of tight-binding systems.
Each site belongs to exactly one \emph{site family}, typically a crystal lattice (monatomic crystals) or a crystal sublattice (polyatomic crystals).
Within a site family, individual sites are distinguished by a \emph{tag}.
In the common case where the site family is a regular lattice, the site tag is simply given by its integer lattice coordinates.

Any crystal lattice with any basis can be created easily by specifying the primitive vectors of its Bravais lattice and the coordinates of its sites inside the lattice unit cell.
For example, the honeycomb lattice of graphene can be created using
\begin{example*}
graphene = kwant.lattice.general(
    [(1, 0), (0.5, 0.5 * sqrt(3))],
    [(0, 0), (0, 1 / sqrt(3))])
\end{example*}
Here, the first argument to \texttt{kwant.lattice.general} is a list of primitive vectors of its Bravais lattice $\vec u_0=(1, 0)$ and $\vec u_1=(1/2,\sqrt{3}/2 )$, and the second is a list of coordinates of the sites inside the unit cell $\vec v_\mathrm{A}=(0, 0)$ (the site of sublattice A) and $\vec v_\mathrm{B}=(0,1/\sqrt{3})$ (the site of sublattice B).
A site family is associated with each sublattice,
\begin{example*}
A, B = graphene.sublattices
\end{example*}
which provides a unique mapping between the position on the lattice and the sites.
For instance, the atom at position $\vec R = 5 \vec u_0 + 8 \vec u_1 + \vec v_\mathrm{B}$ corresponds to the site \texttt{B(5,8)} which belongs to the site family \texttt{B} and has the tag \texttt{(5,8)}.
The position $\vec R$ of a site in real space can be accessed using the property \texttt{site.pos}.

Kwant comes with several common lattices predefined, so that instead of defining the honeycomb lattice as above, one may just use
\begin{example*}
graphene = kwant.lattice.honeycomb()
\end{example*}

In a completely similar fashion, a 3-d cubic lattice may be defined as
\begin{example*}
cubic = kwant.lattice.general([[1, 0, 0],
                               [0, 1, 0],
                               [0, 0, 1]])
\end{example*}
Even though most examples in this article are 2-dimensional, Kwant is not limited to any specific dimensionality and uses a dimensionality-independent graph representation of tight-binding systems.

Site families are in principle more general then Bravais lattices.
For example, one could create a site family for an amorphous material or even label sites with names consisting of characters.
In practice, however, Bravais lattices are sufficient to construct most systems.

\subsection{Tight-binding systems as Python mappings}
\label{app:mappings}
Having introduced an important prerequisite, sites, we now proceed to discuss the structure of builders.
The main idea here is to represent a tight-binding system as a mapping from sites and hoppings (=pairs of sites) to the corresponding submatrices of the Hamiltonian.
Similar to any other mapping in Python, builder items are set using the syntax \texttt{sys[key] = value}.
The following example shows how to create a simple system by setting its sites and hoppings one by one.

First, a square lattice and an empty builder are initialized.
\begin{example}
lat = kwant.lattice.square()
sys = kwant.Builder()
\end{example}

In the next step we add three sites to the system and assign a scalar on-site energy of $1.5$ to each of them.
When assigned as builder values, scalars are interpreted as $1 \times 1$ matrices.
\begin{example}
sys[lat(0, 0)] = 1.5
sys[lat(1, 0)] = 1.5
sys[lat(0, 1)] = 1.5
\end{example}

\begin{figure}
  \centering
  \includegraphics[width=\linewidth]{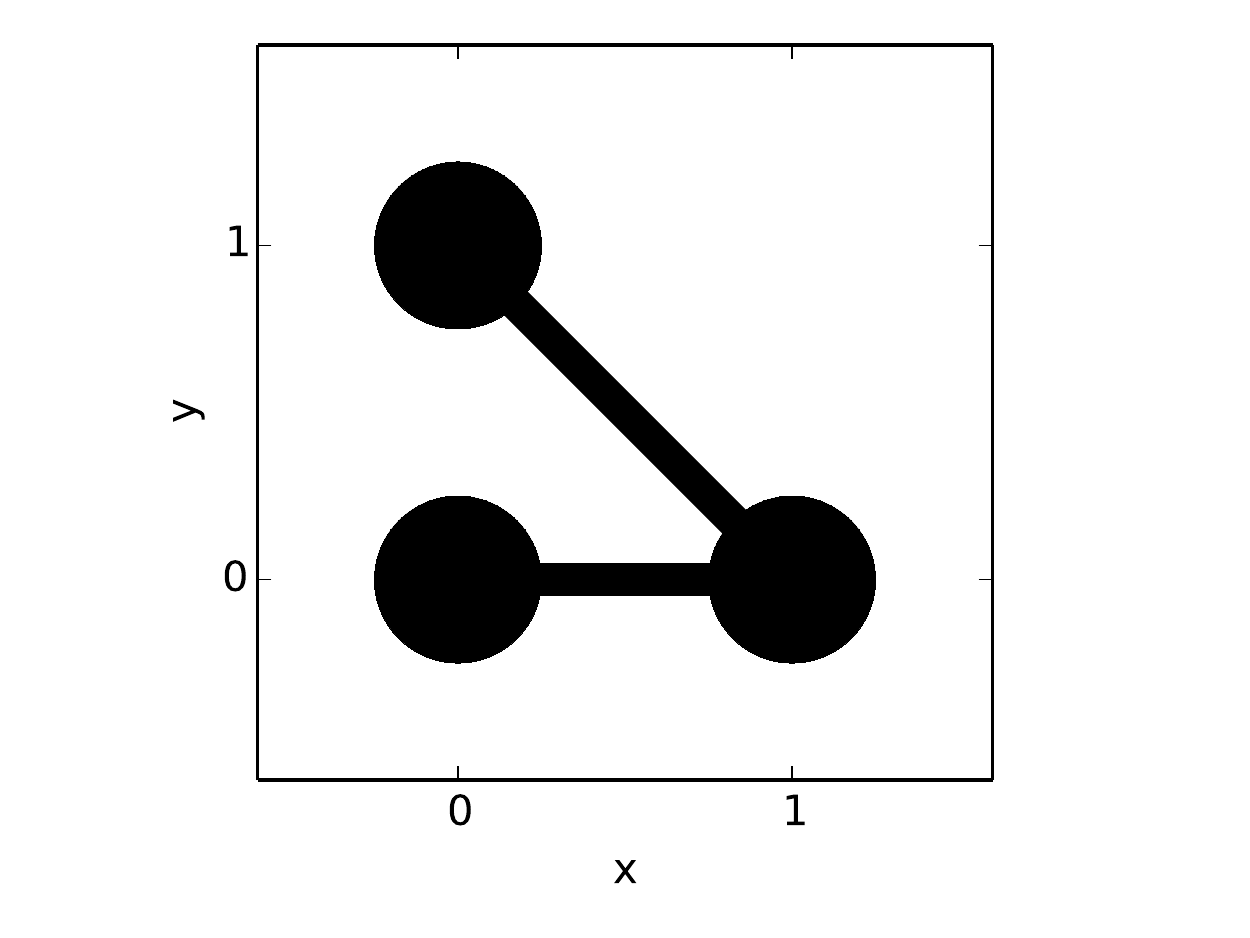}
  \caption{Plot of a very simple tight-binding system consisting of three sites and two hoppings.
    This figure has been generated by the Kwant script shown in Appendix~\ref{ex:simplest}.}
  \label{fig:simplest}
\end{figure}

Finally we add two hoppings to the system.
The syntax now becomes \texttt{sys[site\_to, site\_from] = value}, which sets the value of the hopping from \texttt{site\_from} to \texttt{site\_to}.
\begin{example}
sys[lat(0, 0), lat(1, 0)] = 2j
sys[lat(0, 1), lat(1, 0)] = 2j
\end{example}
The resulting system is shown in Fig.~\ref{fig:simplest}.

Builders are fully-fledged Python mappings.
This means that in addition to setting values of sites and hoppings like in the above example, it is possible to query values, e.g.\ \texttt{print~sys[lat(1, 0)]}.
It is also possible to delete items: \texttt{del~sys[lat(0, 1)]}.

Builder objects ensure their consistency during manipulation:
They automatically remain Hermitian, and hoppings may only exist between sites present in the builder.
Hence, the value of each hopping $(i, j)$ is invariably bound to be the Hermitian conjugate of the hopping $(j, i)$, and when a site is deleted, all of the hoppings to it and from it are deleted as well.

\subsection{Values of sites and hoppings}
\label{app:values}
In the previous subsection the onsite Hamiltonians and hopping values were just scalar constants.
While this allows to define any tight-binding system by introducing a sufficient number of sublattices, in many cases it is useful to also allow these values to be matrices.
Thanks to the flexibility of Python, this works in the most natural way in Kwant.
If instead of a scalar value of the Hamiltonian one needs to use a matrix value, one may just write
\begin{example*}
sys[lat(0, 0)] = numpy.array([[0, -1j], [1j, 0]])
sys[lat(0, 1)] = tinyarray.array([[1, 0], [0, -1]])
\end{example*}
Here we set the Hamiltonian of two sites to be equal to the Pauli matrices $\sigma_y$ and $\sigma_z$.
We used two different ways to define a matrix-like object: as a NumPy array, and a tinyarray.
NumPy \cite{oliphant07} is the standard Python array library, and tinyarray is a library developed for Kwant, specifically optimized to be fast with small arrays, as discussed in Appendix~\ref{app:implementation}.
In this application, the only difference between these two is performance, which is better for tinyarray arrays.
The number of orbitals may be different for different sites as long as the sizes of all the value matrices are consistent:
the hopping integral $H_{ij}$ from site $j$ to $i$ must have the size $n_i \times n_j$, with $n_i$ and $n_j$ the sizes of onsite Hamiltonian of these two sites.
(In practice all the Hamiltonian element matrices are very often square and of the same size, e.g.\ $2 \times 2$ for a model with spin.)

In order to simplify the definition of more complicated Hamiltonians, Kwant also supports values that are functions.
If a function is assigned as a value of a builder for some site or hopping, its evaluation is postponed until the last possible moment, that is when the Hamiltonian is actually used in a calculation.
(Performing these calculations is the topic of Appendix~\ref{app:solvers}.)
Setting functions as values with a builder works exactly like setting constant values: \texttt{sys[key] = value}, with the only difference that the \texttt{value} is now a function, not a number or a matrix.
The value function is called with the corresponding site or hopping as arguments, and optionally some additional parameters.

Value functions help to elegantly define numerical values that depend on position.
Furthermore, they allow to use different numerical values with a single finalized system (varying parameters such as magnetic field or electrostatic potential).

\subsubsection{Example: value functions for hoppings and sites}
\label{app:values_ex_qhe}
As an example let us use value functions to model a system with constant perpendicular magnetic field and disorder.
The following code snippets are part of the quantum Hall effect example shown fully in Appendix~\ref{ex:qhe} and further used in Appendix~\ref{app:transport}.

First, we consider the hoppings.
We choose the vector potential in the Landau gauge with $\bm{B} = B \hat{z}$, so that $\bm{A} = -By \hat{x}$.
Including a vector potential in a tight-binding system is done using Peierls substitution \cite{peierls33,hofstadter76}.
The hopping integral $t_{ij}(\Phi)$ from the point $(x_j, y_j)$ to the point $(x_i, y_i)$ is then given by
\begin{equation}
  \label{eq:peierls}
  t_{ij}(\Phi) = t_{ij}(0) \times \re^{-\ri \Phi (x_i - x_j) (y_i + y_j)/2},
\end{equation}
with $\Phi = B e a^2 / \hbar$ the flux per lattice unit cell in units of flux quanta.
Note that we have chosen $\mathbf{A}$ such that $t_{ij}$ does not depend on $x$.
This will allow us to use the same gauge in $x$-directed leads.\footnote{
  It is also possible to add a magnetic field to a system with non-parallel leads \cite{baranger89}: For each lead, an adopted Landau gauge is chosen.
  For the scattering region, a Landau gauge is also adopted with local gauge transformations applied to the interface sites of each lead that mediate between the different Landau gauges.
  A Kwant module that automates this procedure is planned to be made available.}
$t_{ij}$ can be defined as a value function in Kwant as
\begin{example}
def hopping(sitei, sitej, phi, salt):
    xi, yi = sitei.pos
    xj, yj = sitej.pos
    return -exp(-0.5j * phi * (xi - xj) * (yi + yj))
\end{example}
This function receives four arguments:
The first two are the sites that are connected by the hopping for which the value is requested.
The remaining two are user-specified additional parameters.
\texttt{phi} corresponds to $\Phi$.
\texttt{salt} is not used here, but it will be used in the function \texttt{onsite} below.
Declaring that second parameter is necessary because each value function of a given system receives the same user-specified parameters.

Having specified the hoppings, we define an uncorrelated Gaussian disorder potential for the sites.
The usual approach for defining disorder is to use a random number generator and evaluate the disordered potential on each site in sequence.
With Kwant, however, solvers are allowed to evaluate the value of a site or hopping more than once and expect that it does not change during a single invocation of a solver.
Additionally, since the order in which the values of sites and hoppings are evaluated is undefined and depends on internal details of the builder and the used solver.
One solution to this problem would be to generate a large table of random numbers in advance and to look up the potential for each site in this table.
Kwant offers another solution, that is easier to handle especially for non-square lattices: a random-access pseudo random number generator provided by the module \texttt{kwant.digest}\footnote{
The module \texttt{kwant.digest} provides routines that given some input compute a ``random'' output that depends on the input in a (cryptographically) intractable way \cite{tzeng08}.
This turns out to be very useful when one needs to map some irregular objects to random numbers in a reproducible way.
Internally, the MD5 hash algorithm \cite{rivest92} is used.
The randomness generated in this fashion is good enough to pass the ``dieharder'' \cite{brown13} battery of randomness tests.
}.
Defining an uncorrelated Gaussian disorder using this generator amounts to the following value function:
\begin{example}
def onsite(site, phi, salt):
    return 0.05 * gauss(repr(site), salt) + 4
\end{example}
Since this is a value function for sites, the first argument has to be the site on which it is going to be evaluated.
\texttt{phi} is the user-specified parameter that was used with \texttt{hopping} above and is ignored here.
Passing different string values to the \texttt{salt} parameter results in different realizations of the disorder, so \texttt{salt} plays a role similar to a random seed.



\subsubsection{Example: value functions that return matrices}
\label{app:values_ex_majorana}
The following second example, quite different from the preceding one, demonstrates the flexibility of value functions.
We consider a section of the Majorana fermion script of Appendix~\ref{ex:majorana}.
It first defines the Pauli matrices
\begin{example}
s_0 = numpy.identity(2)
s_z = numpy.array([[1, 0], [0, -1]])
s_x = numpy.array([[0, 1], [1, 0]])
s_y = numpy.array([[0, -1j], [1j, 0]])
\end{example}
and some Kronecker-products of them
\begin{example}
tau_z = tinyarray.array(numpy.kron(s_z, s_0))
tau_x = tinyarray.array(numpy.kron(s_x, s_0))
sigma_z = tinyarray.array(numpy.kron(s_0, s_z))
tau_zsigma_x = tinyarray.array(numpy.kron(s_z, s_x))
\end{example}
The calls to \texttt{tinyarray.array} are optional.
Their purpose is to improve the performance of the value functions.
These constants are used within the value functions that define the site Hamiltonians
\begin{example}
def onsite(site, p):
    return tau_z * (p.mu - 2 * p.t) + \
        sigma_z * p.B + tau_x * p.Delta
\end{example}
and the hopping integrals
\begin{example}
def hopping(site0, site1, p):
    return tau_z * p.t + 1j * tau_zsigma_x * p.alpha
\end{example}
Note the following features of these value functions:
\begin{itemize}
  \item Their return values are matrices.
  A value function may return anything that would be valid as a constant value for a builder.
  \item They do not depend on their site parameters.
  Still, Kwant will call them for each site and hopping individually such that the same values will be re-calculated many times.
  Even though this may seem wasteful, as discussed in Appendix~\ref{app:implementation} such inefficiencies do not play a role in practice.
  \item A single extra parameter \texttt{p} is passed, as opposed to the many parameters that define the Hamiltonian.
  \texttt{p} is a namespace object (see Appendix~\ref{ex:majorana} for its definition) that contains all the actual parameters as its attributes.
  This technique is useful to avoid mistakes when passing many parameters to value functions.
\end{itemize}

\subsection{Acting on multiple sites/hoppings at once}
\label{app:iterables}
The features of builders that have been introduced so far are in principle sufficient to define (site-by-site and hopping-by-hopping) systems with an arbitrarily complex geometry.
In order to facilitate a higher level of abstraction, in addition to the simple keys (single sites and single hoppings) builders can also be indexed by composite keys that combine multiple simple keys.
This feature has been inspired by the ``slicing'' of some Python containers and by NumPy's ``fancy indexing''.

The most basic kind of these advanced keys is a list of simple keys.
With such a list it is possible to assign a value to multiple sites or hoppings in one go.
We now create a list of sites of the lattice \texttt{lat} belonging to a disk.
\begin{example*}
sites = []
for x in range(-10, 11):
    for y in range(-10, 11):
        if x**2 + y**2 < 13**2:
            sites.append(lat(x, y))
\end{example*}
This list of sites can be added to the system in one step:
\begin{example*}
sys[sites] = 0.5
\end{example*}

Any non-tuple object that supports iteration can be used just like a list.
For example, the following code snippet adds the same sites using a \emph{generator expression}.
The syntax is quite self-explanatory.
We refer to the Python tutorial \cite{python_tutorial_classes} for further information.
\begin{example*}
sys[(lat(x, y)
     for x in range(-10, 11)
     for y in range(-10, 11)
     if x**2 + y**2 < 13**2)] = 0.5
\end{example*}

Finally, the most high-level way to add many sites at once is to use a method of the lattice named \texttt{shape}.
This method finds all the sites of a lattice that fit inside a certain region and can be reached from a given starting point.
Having defined a function that specifies the circular region
\begin{example}
def disk(pos):
    x, y = pos
    return x**2 + y**2 < 13**2
\end{example}
the code required for adding the same sites as before to the system is very compact:
\begin{example}
sys[lat.shape(disk, (0, 0))] = 0.5
\end{example}
Note that we do not have to explicitly loop over the possible candidate sites anymore.
The advantage of using \texttt{shape} becomes greater for non-square lattices and for more complicated shapes.
We will see an example of such usage further below.

With a Kwant builder, one will often first add all the sites to a system, and then all the hoppings that connect them.
If we regard all hoppings that only differ by a lattice translation as a single kind, that second step typically adds only a few kinds of hoppings.
These are represented by objects of the type \texttt{HoppingKind} from the module \texttt{kwant.builder} that can be used directly as keys:
\texttt{HoppingKind(displacement, lattice2, lattice1)} generates all the hoppings inside the system that start at \texttt{lattice1}, end at \texttt{lattice2}, and whose lattice coordinates differ by \texttt{displacement}.
Using \texttt{HoppingKind} to set all the nearest neighbor hoppings of a system on a square lattice can be done as
\begin{example*}
sys[kwant.builder.HoppingKind((1, 0), lat, lat)] = 1
sys[kwant.builder.HoppingKind((0, 1), lat, lat)] = 1
\end{example*}
To go even further, Kwant can calculate a list of all the \mbox{n-th} nearest neighbor hopping kinds on any lattice.
That list, returned by the method \texttt{neighbors} can be used directly as a key:
\begin{example}
sys[lat.neighbors(1)] = 1
\end{example}
(The argument 1 to \texttt{lat.neighbors} means that nearest neighbors are to be found and could have been omitted since it is the default value.)
This high-level key allows to access the n-th nearest neighbors of any regular lattice in a single line of code.
The complete yet very compact construction script for the circular quantum dot is listed in Appendix~\ref{ex:disk} and its output is shown in Fig.~\ref{fig:disk}.

\begin{figure}
  \centering
  \includegraphics[width=\linewidth]{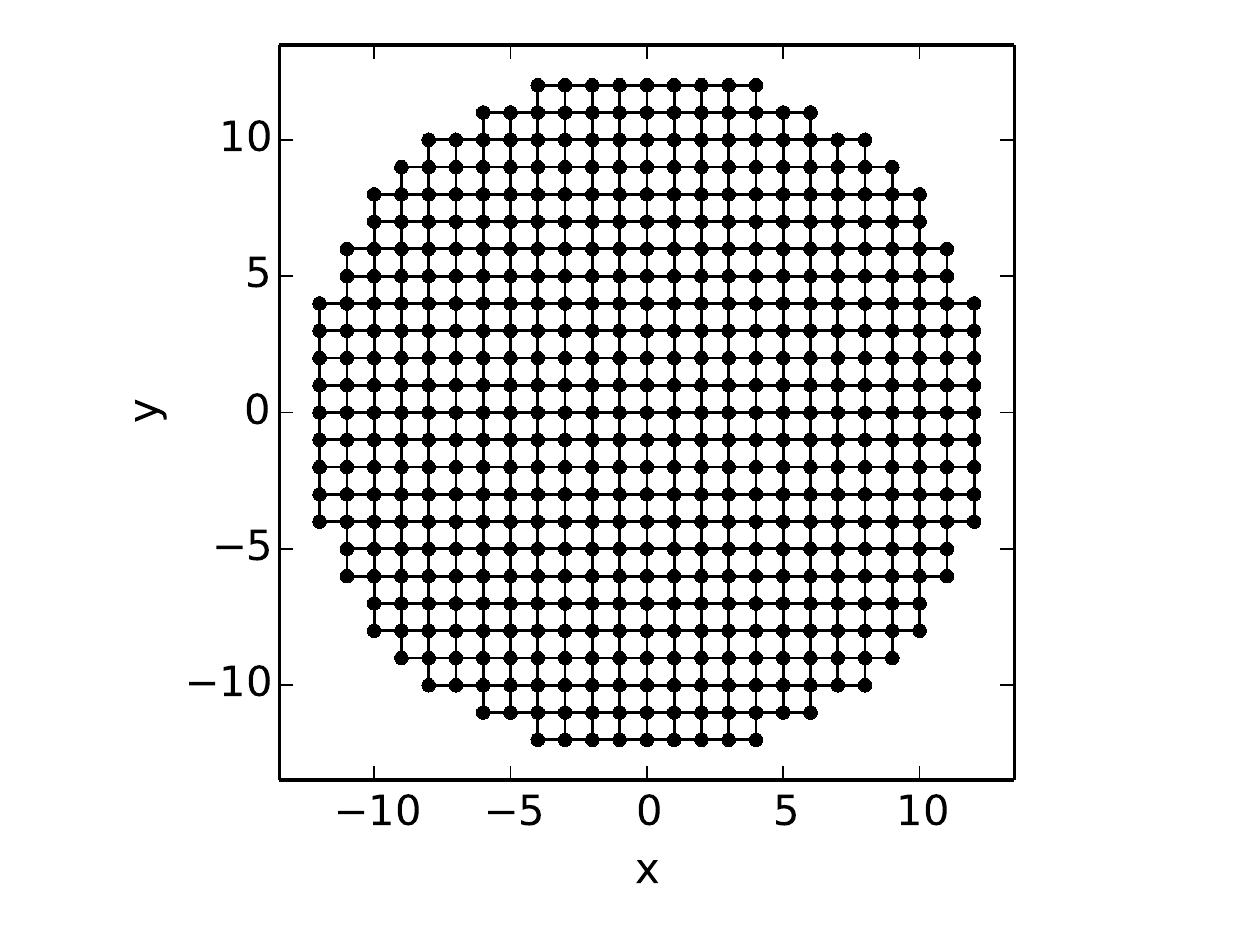}
  \caption{Plot of a circular quantum dot.
    This figure has been generated by the Kwant script shown in Appendix~\ref{ex:disk}.}
  \label{fig:disk}
\end{figure}

For further information about keys we refer to the documentation of Kwant, specifically the documentation of the method \texttt{expand} of \texttt{Builder}.
Here, we will just show another example that demonstrates the flexibility of the approach.
Consider Fig.~\ref{fig:bean} that has been generated by the script shown in Appendix~\ref{ex:bean}.
This program is identical to the one for the disk except that it features a different type of lattice
\begin{example}
lat = kwant.lattice.honeycomb()
\end{example}
and a different region-defining function:
\begin{example}
def bean(pos):
    x, y = pos
    rr = x**2 + y**2
    return rr**2 < 15 * y * rr + x**2 * y**2
\end{example}
Note that the function defining the shape (\texttt{bean} in the example) receives points in real space.
In Kwant, each site ``lives'' in two coordinate systems: the coordinates system of the lattice (with integer coefficients) and the real-space ``world'' coordinates.
These two coordinate systems are only identical in the simplest square lattice (that is actually used in several examples of this article), but not in general, like in the example above.

\begin{figure}
  \centering
  \includegraphics[width=\linewidth]{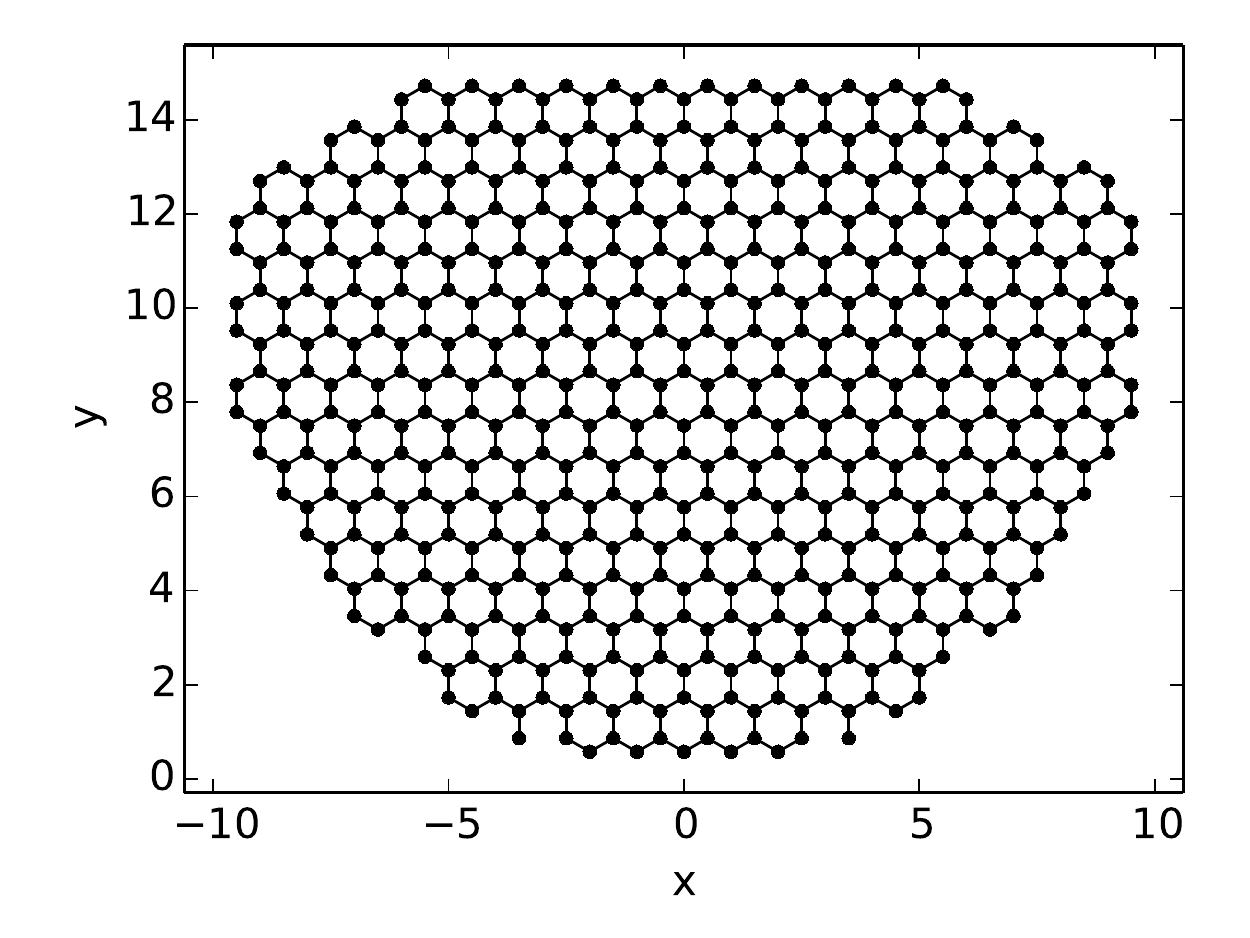}
  \caption{Plot of an irregularly-shaped graphene quantum dot.
    Note that thanks to Kwant's high-level abstractions of lattices and shapes, the script that generates this figure (Appendix~\ref{ex:bean}) is very similar to the one for Fig.~\ref{fig:disk}.}
  \label{fig:bean}
\end{figure}

\subsection{Symmetries}
\label{app:symmetries}
So far all examples in this section involved systems with a finite number of sites.
Kwant supports infinite periodic systems as well:
Upon creation of a builder a spatial symmetry can be specified that will be automatically enforced for that system.
Infinite systems with a translation symmetry can be attached to finite systems as leads.

Kwant's abstraction of symmetries closely matches the mathematical concept of spatial symmetry:
A symmetry in Kwant is specified by a set of spatial transformations of sites and hoppings that generate the symmetry group
and a set of sites and hoppings called \emph{fundamental domain} that contains a single representative from each set of mutually symmetrical sites/hoppings\footnote{For hoppings, the fundamental domain is defined as follows:
A hopping $(i, j)$ belongs to the fundamental domain exactly when the site $i$ belongs to the fundamental domain.}.
In order to completely describe an infinite periodic tight-binding system it is thus sufficient to know the symmetry and those sites and hoppings of the system that lie within its fundamental domain.

In practice, users of Kwant never specify symmetries in this low-level way, they use a predefined symmetry class instead.
(Translational symmetry is by far the most important for defining scattering problems, and this is why it is the only class predefined in Kwant.)
A \texttt{kwant.TranslationalSymmetry} object is initialized by one or several real-space vectors, each representing a translation under which the system is to remain invariant.
Currently, the user has no control over the fundamental domain -- it is deduced from the translation symmetry periods using a simple algorithm.

Sites and hoppings can be added and manipulated in the same way as in finite systems, with the difference that each site/hopping is now treated as a representative of all the sites/hoppings symmetrical to it.
The following example first creates a builder with a symmetry.
Then, all the sites symmetrical to \texttt{lat(3, 0)} are added.
Finally, the sites that have been just added are deleted using a site as key that differs from \texttt{lat(3, 0)} but is symmetrical to it.
\begin{example*}
sym = kwant.TranslationalSymmetry((2, 0))
sys = kwant.Builder(sym)
sys[lat(3, 0)] = 0
del sys[lat(-1, 0)]
\end{example*}

To achieve this behavior, a builder with a symmetry internally maps all sites and hoppings to the fundamental domain.
Each serves as a unique representative for itself and all of its images under the symmetry.
This mapping is often invisible to the user, but can lead to confusion when the sites or hoppings of a builder with a symmetry are printed and seem to differ from the ones that were set.

Due to limitations of the current low-level system format, in Kwant 1.0 it is only possible to finalize systems with 1-d translational symmetry.

\subsection{Leads}
\label{app:leads}
Infinite periodic systems can be attached to finite systems as \emph{leads}.
Kwant uses the convention that the period of the lead symmetry must point away from the scattering region.
Attaching a lead works by specifying a set of sites of the finite system to which the lead is to be attached: the \emph{lead interface}.
The sites of the lead interface must belong to an image of a single unit cell of the lead under the lead symmetry, and the hopping between the lead and the lead interface is assumed to be equal to the hopping between the neighboring lead unit cells.
This can always be achieved, requiring sometimes to add an extra unit cell of the lead to the scattering region.

To make attaching of leads easier, builders provide the method \texttt{attach\_lead}.
This method automatically adds sites and hoppings to the scattering region to make it compatible with the lead shape and the requirements imposed on the lead interface.
The added sites and hoppings are assigned the values of the corresponding sites and hoppings in the lead.
The lead interface is then calculated automatically.

The following example (with complete code shown in Appendix~\ref{ex:leads}) creates a ring-shaped scattering region.
It then defines an infinite system with period $(-2, 1)$ and adds to it a horizontal line of 12 sites with coordinates ranging from $(-6, 0)$ to $(5, 0)$.
Thanks to the symmetry mechanism, the infinite system contains also all images of that horizontal line under all multiples of the period.
This infinite system is attached twice as two different leads to the scattering region: once outside and once inside.
\begin{example}
def ring(pos):
    x, y = pos
    return 7**2 <= x**2 + y**2 < 13**2

sys[lat.shape(ring, (10, 0))] = 0
sys[lat.neighbors()] = 1

sym = kwant.TranslationalSymmetry((-2, 1))
lead = kwant.Builder(sym)
lead[(lat(x, 0) for x in range(-6, 6))] = 0
lead[lat.neighbors()] = 1
sys.attach_lead(lead)
sys.attach_lead(lead, lat(0, 0))
\end{example}
The argument \texttt{lat(0, 0)} to the second call of \texttt{attach\_lead} specifies the starting point for the algorithm: the lead is attached at the nearest possible place in the direction opposite to the lead direction.
As they are attached, leads are labeled with consecutive integers starting from zero.

\begin{figure}
  \centering
  \includegraphics[width=\linewidth]{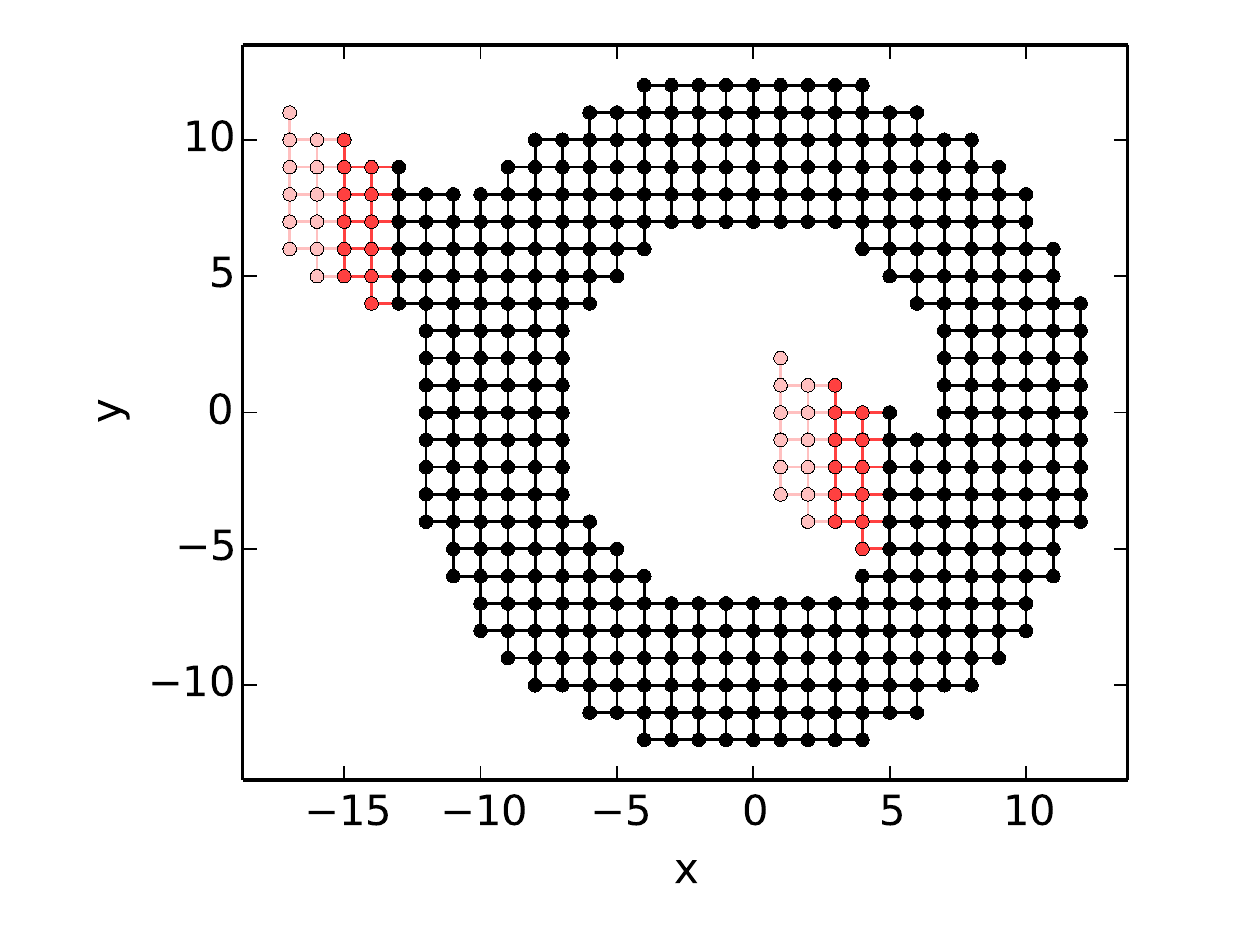}
  \caption{Plot of a ring-shaped scattering region (black dots) with two identical leads (red dots).
    Each semi-infinite periodic lead is represented by its first two unit cells which are shown in different shades of red.
    Note that all unit cells have the same shape and that sites were added to the scattering region such that the leads are connected well.
    This figure has been generated by the Kwant script shown in Appendix~\ref{ex:leads}.}
  \label{fig:leads}
\end{figure}

A plot of the resulting system is shown in Fig.~\ref{fig:leads}.
The scattering region is represented by the black dots.
The lead is represented by its first two unit cells (colored dots).
Note that the lead unit cells are not perpendicular to the lead direction, as their shape is determined by the fundamental domain and the latter is chosen implicitly.
This does not affect the physics of the system, it does, however, affect which sites have to be added to the system by \texttt{attach\_lead}.

\section{Calculations with tight-binding systems}
\label{app:solvers}

\subsection{Low-level systems}
\label{app:system}
\begin{figure}
\begin{center}
\includegraphics[width=0.9\linewidth]{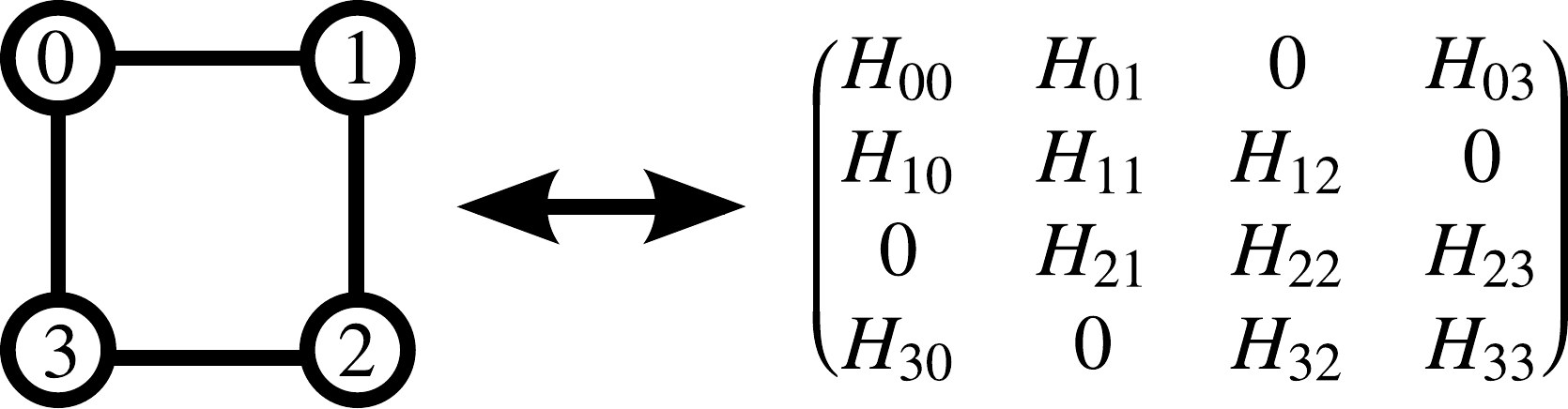}
\end{center}
\caption{In the low-level system, a graph (left) is used to represent a Hamiltonian matrix (right).}
\label{fig:graph}
\end{figure}

Since tight-binding Hamiltonians are sparse, it is natural to store them as annotated graphs.
That representation of the Hamiltonian matrix is illustrated in Fig.~\ref{fig:graph}: A tight-binding
degree of freedom $i$ corresponds to a node of a graph, every non-zero hopping matrix element $H_{ij}$ to an edge connecting nodes $i$ and $j$.
The graph thus represents the structure of the non-zero entries of the Hamiltonian.
Together with the values $H_{ij}$ the complete tight-binding Hamiltonian is defined.

Such annotated graphs are called low-level systems within Kwant and are implemented by a hierarchy of classes of objects in the module \texttt{kwant.system}.
As explained in Sec.~\ref{sec:foundations}, low-level systems form the common interface between the two phases of construction and solving.
Within a low-level system, the graph structure itself is stored in a compressed sparse row format \cite{barrett94}.
The values of the onsite Hamiltonians and hoppings $H_{ij}$ are provided upon request by the method \texttt{hamiltonian} given $i$, $j$ and possibly other parameters that are required to evaluate the Hamiltonian matrix elements (see Appendix~\ref{app:values}).
Periodic infinite systems that can be attached as leads to a finite system are represented in a similar way.
Since no Python-specific features are used, low-level systems are compatible with C, Fortran and similar languages.
This makes Kwant independent of a single programming language: even though the current version of Kwant is mostly written in Python, systems and tools for working with systems can be implemented in other programming languages as well.

Most often, a low-level system will be obtained from a \texttt{Builder} instance by calling the method \texttt{finalized}.
In this case, the mapping from system sites to the integers numbering the graph is not controlled by the user.

It is also perfectly possible to implement a low-level system directly, deriving from \texttt{FiniteSystem}:
\begin{example}
class SquareMolecule(kwant.system.FiniteSystem):
    def __init__(self):
        g = kwant.graph.Graph()
        g.add_edges([(0, 1), (1, 0),
                     (1, 2), (2, 1),
                     (2, 3), (3, 2),
                     (0, 3), (3, 0)])

        self.graph = g.compressed()
        self.leads = self.lead_interfaces = []

    def hamiltonian(self, i, j, E=0.1, t=1):
        return E if i == j else t
\end{example}
The Hamiltonian of this system is printed by
\begin{example}
dm = SquareMolecule()
print dm.hamiltonian_submatrix().real
\end{example}
which outputs
\begin{example*}
[[ 0.1  1.   0.   1. ]
 [ 1.   0.1  1.   0. ]
 [ 0.   1.   0.1  1. ]
 [ 1.   0.   1.   0.1]]
\end{example*}
revealing the matrix structure of the example shown in Fig.~\ref{fig:graph}.
Here we have used the convenience method \texttt{hamiltonian\_submatrix} that can return any submatrix of the Hamiltonian as a NumPy array, or as a SciPy sparse matrix, and that returns the full Hamiltonian matrix by default.

The numbering of the graph and the numbering of the indices of the entries in the full Hamiltonian matrix coincide when the hopping matrix elements $H_{ij}$ are all scalar.
In general, however, the hopping matrix elements can be $n_i \times n_j$ matrices $H_{ij}$.
The Hamiltonian matrix should then be interpreted as a block matrix, with the graph indices as block indices.

\subsection{Quantum transport}
\label{app:transport}
Given the Hamiltonian matrix of the system and of the leads attached to it, various methods may be used to compute transport properties.
The default algorithm used in Kwant is the wave function approach that has been introduced in Sec.~\ref{sec:scattering_theory} and amounts to setting-up and solving of a sparse system of linear equations.
The full algorithm, including the case of non-invertible hopping matrices, a numerically stabilized version of the algorithm and a discussion of the connection to the Green's function formalism will be published elsewhere \cite{wimmer13}.

The sparse matrix of the full linear system to be solved contains the full Hamiltonian of the scattering region and additional rows and columns for outgoing and evanescent modes of the leads.
At first glance, a direct solution might seem inefficient and even infeasible for large systems.
However, there is a large variety of established libraries for solving general sparse systems of linear equations that allow nevertheless for a very efficient and stable solution.
The efficiency of these sparse linear solvers depends crucially on choosing a good ordering of the coefficient matrix.
For systems arising from regular grids, nested dissection orderings have been shown to be optimal.
For such orderings, the time needed for solving a two-dimensional tight-binding system with the geometry of a rectangle of length $L$ and width $W$ scales as $O(L W^2)$ \cite{George73}.\footnote{
  This scaling is derived under the assumption that no pivoting needs to be done.
  With pivoting, the scaling can be worse.
  However, pivoting guarantees a stable solution, and in practical applications we have observed that our algorithm could stably solve problems for which RGF failed.}
This is a more favorable scaling compared to the RGF algorithm whose execution time scales as $O(L W^3)$ \cite{lee1981,thouless1981,mackinnon1985}.
Indeed, in practice we find that the transport algorithm in Kwant is considerably faster than RGF for systems of intermediate and large size, as shown in Sec.~\ref{sec:benchmark}.

\begin{figure}
  \centering
  \includegraphics[width=\linewidth]{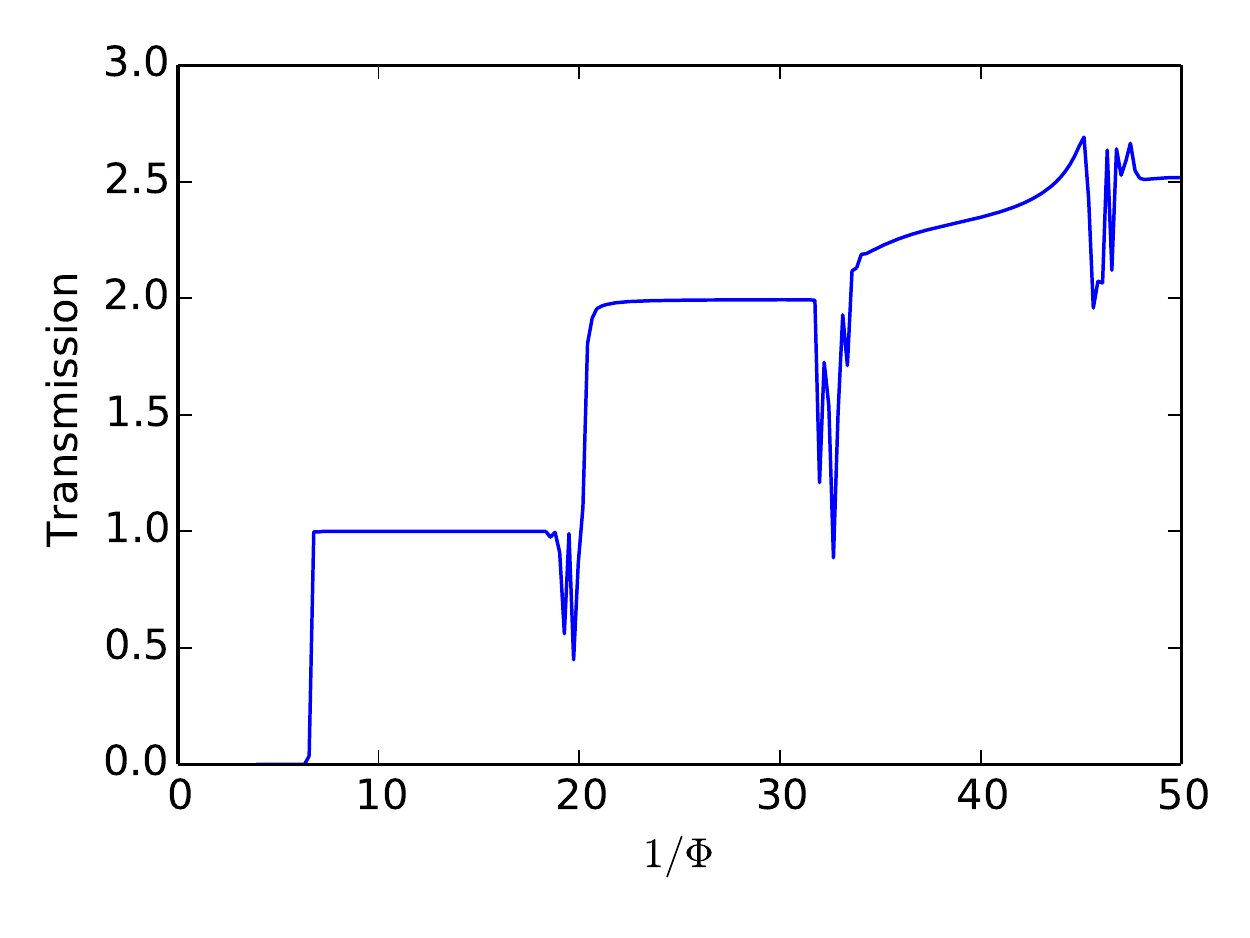}
  \caption{Quantum Hall effect conductance plateaus in the presence of disorder.
    The first two plateaus show the quantization of conductance that is the hallmark of the quantum Hall effect.
    The third plateau around $1 / \Phi = 40$ does not develop due to a constriction in the system that leads to backscattering.
    The situation at this value of magnetic field is shown in Fig.~\ref{fig:qhe_edgestate}.
    This figure has been generated by the Kwant script shown in Appendix~\ref{ex:qhe}.}
  \label{fig:qhe_plateaus}
\end{figure}

Solving the linear system of equations yields the scattering matrix.
Within the framework of Kwant, the function \texttt{kwant.smatrix} performs this calculation given a low-level system, the Fermi energy and optional user-defined parameters of the system.
It returns a scattering matrix object that can be queried for e.g.\ the values of transmission or noise between leads.
The following code, for example, calculates conductance as a function of a range of magnetic fluxes.
Together with the value functions of Appendix~\ref{app:values_ex_qhe} and a few lines of code that construct the system that are shown in the full listing in Appendix~\ref{ex:qhe}, the quantum Hall effect conductance plateaus of Fig.~\ref{fig:qhe_plateaus} are generated.
Note that the list passed as the \texttt{args} parameter contains the values of the two user-defined parameters of the value functions.
The arguments to \texttt{transmission} specify that the conductance from lead 0 to lead 1 is to be returned.
\begin{example}
reciprocal_phis = numpy.linspace(4, 50, 200)
conductances = []
for phi in 1 / reciprocal_phis:
    smatrix = kwant.smatrix(sys, energy, args=[phi, ""])
    conductances.append(smatrix.transmission(1, 0))
\end{example}

\begin{figure}
  \centering
  \includegraphics[width=\linewidth]{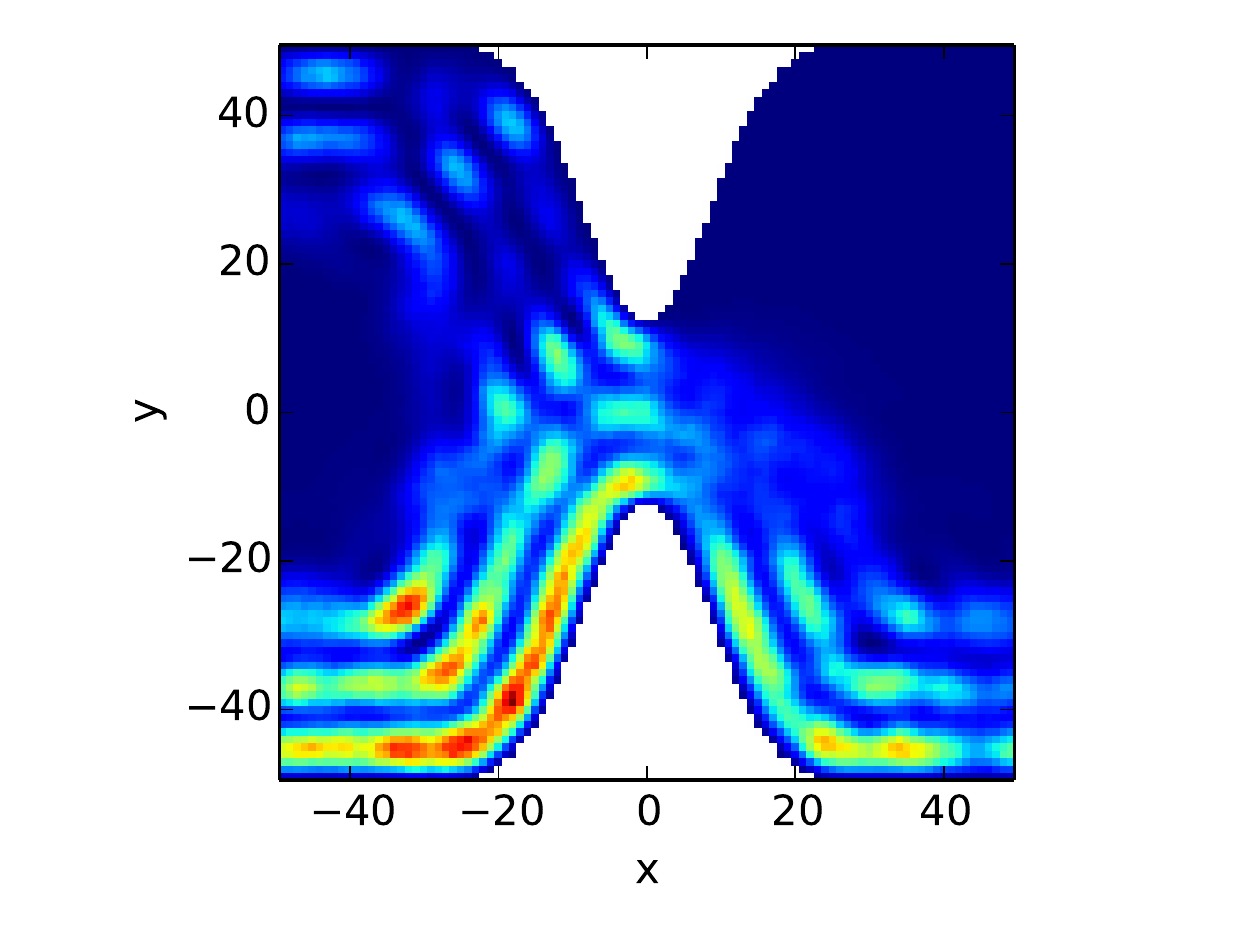}
  \caption{Density of a quantum Hall edge state that is partially backscattered at $1 / \Phi = 40$ due to a constriction.
    This figure has been generated by the Kwant script shown in Appendix~\ref{ex:qhe}.}
  \label{fig:qhe_edgestate}
\end{figure}

In fact, \texttt{kwant.smatrix} is a shortcut for the function \texttt{smatrix} of \texttt{kwant.solvers.default}, the default solver module.
This module offers additional functionality as well, for example the calculation of the scattering wave function \eqref{eq:scat_wave_func2} of each scattering state \eqref{eq:scat_wave_func}.
When \texttt{kwant.smatrix} is used, these wave functions are discarded since keeping them may require an excessive amount of memory for large systems.
Instead, when it is needed, the wave function can be obtained using \texttt{kwant.wave\_function}.
This routine prepares another function that must be supplied the lead number as the only argument to finally get an array of the scattering wave functions corresponding to the incoming modes of that lead.
The following code uses this mechanism to calculate the local density of all the states incoming from a given lead.
\begin{example}
def density(sys, energy, args, lead_nr):
    wf = kwant.wave_function(sys, energy, args)
    return (abs(wf(lead_nr))**2).sum(axis=0)
\end{example}
It returns an array that contains a density value for each site of the system.
Such an array can be color-plotted with the function \texttt{map} of Kwant's plotter module:
\begin{example}
d = density(sys, energy, [1/40.0, ""], 0)
kwant.plotter.map(sys, d)
\end{example}
The result, shown in Fig.~\ref{fig:qhe_edgestate}, shows the backscattering of the quantum Hall edge state at $1 / \Phi = 40$ that is the reason for the absence of the third conductance plateau in Fig.~\ref{fig:qhe_plateaus}.

\subsection{Exact diagonalization of a finite system Hamiltonian}
\label{app:diag}
One of the design objectives of Kwant was to make it easy to perform arbitrary user-specified computations with complex tight-binding systems.
When the included solvers do not provide the desired calculation it can be often implemented in just a few lines of Python.
(In fact, this is nothing else than writing a solver.)

As an example, we apply the ARPACK library \cite{lehoucq97} to solve a large scale eigenvalue problem and calculate the lowest eigenenergies of a finite system.
ARPACK is easily accessible from Python thanks to SciPy.
Kwant's \texttt{hamiltonian\_submatrix} can return the Hamiltonian matrix in a sparse format understood by SciPy, so there remains almost nothing to be done:
\begin{example*}
H = sys.hamiltonian_submatrix(sparse=True)
print scipy.sparse.linalg.eigsh(H, k=20, which='SM')[1]
\end{example*}

\begin{figure}
  \centering
  \includegraphics[width=\linewidth]{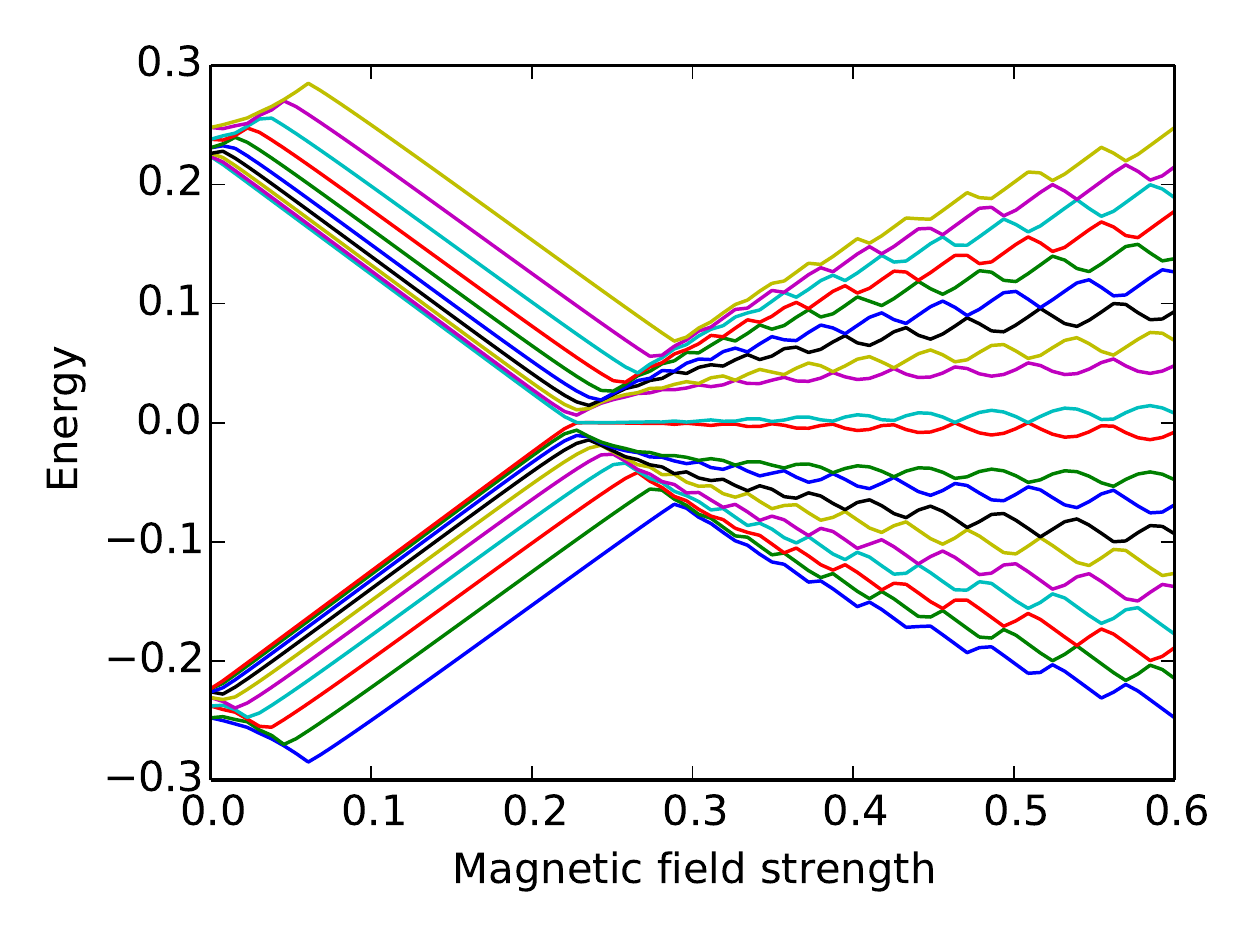}
  \caption{Example of sparse eigenvalue calculations.
    Energy of 20 levels with lowest excitation energies in a Majorana wire, as a function of magnetic field.
    This figure has been generated by the Kwant script shown in Appendix~\ref{ex:majorana}.}
  \label{fig:majorana}
\end{figure}

Together with the value functions of Appendix~\ref{app:values_ex_majorana} and a few lines of code only shown in Appendix \ref{ex:majorana} this functionality is used in the following code snippet to calculate the energy of the states closest to the Fermi level in a Majorana wire as a function of magnetic field strength:
\begin{example}
B_values = numpy.linspace(0, 0.6, 80)
energies = []
params = SimpleNamespace(
    t=1, mu=-0.1, alpha=0.05, Delta=0.2)
for params.B in B_values:
    H = sys.hamiltonian_submatrix(
        args=[params], sparse=True)
    H = H.tocsc()
    eigs = scipy.sparse.linalg.eigsh(H, k=20, sigma=0)
    energies.append(numpy.sort(eigs[0]))
\end{example}
The output of this code is shown in Fig.~\ref{fig:majorana}.

\section{Implementation of Kwant}
\label{app:implementation}
We have demonstrated that the use of the high-level dynamic language Python for the interface of a quantum transport library can offer considerable advantages in terms of usability.
Since the most straightforward way to create a library with a Python interface is to write it in Python and also due to the expressiveness of this language, we have striven to not only provide a Python interface, but to also use Python as much as possible for the implementation of the library itself.
That, however, brings in the risk of drastically decreased performance compared to the compiled languages used traditionally for similar tasks -- a carelessly written Python program can be about 100 times slower than its C counterpart.
By employing a number of measures that are the focus of this section we have ensured that the performance of Kwant remains competitive with other quantum transport codes.
In fact, due to a novel approach to solving the scattering problem (see Appendix~\ref{app:solvers} and Ref.~\cite{wimmer13}) Kwant can be more than an order of magnitude faster than traditional quantum transport codes for large systems, as shown in Sec.~\ref{sec:benchmark}.

\subsection{Resource usage of quantum transport calculations}
The individual steps of a quantum transport calculation have running times that scale differently with $n$, the number of sites in the system.
The asymptotically most expensive step is the solving: the execution time scales as $O(n^{3 - 2/d})$ for a $d$-dimensional system in the case of the RGF algorithm, for instance.
Most other parts of the calculation such as initialization, definition of the system, preparation of solving, and (typical) post-processing of the results, have an asymptotic execution time between $O(1)$ and $O(n)$.
Similar considerations are valid for memory usage with the memory cost of the solving step dominating as well.

The parts of a quantum transport code with asymptotically highest cost are typically small in terms of code size compared to the rest.
This has the important consequence that one may allow oneself to implement most of the code with less attention to performance without a significant penalty.
This is not merely an excuse to work sloppily:
Freeing oneself from the constraint to strive for the highest possible computer efficiency allows to focus on other aspects like human-time efficiency both of the users and the authors of the code.
This insight is the guiding principle behind the technical choices that were made when implementing Kwant.

\subsection{Programming languages used}
Kwant is written primarily in Python.
While some parts of it are implemented in other languages, all these pieces are held together by Python code.
Care has been taken to optimize all the asymptotically most expensive components.
For the less critical parts pragmatic choices were made:
If some component was measured to be too inefficient for typical work loads, it was optimized gradually until a satisfactory performance was obtained.
Overall, all of the following approaches are present in Kwant:
\begin{enumerate}
  \item Usage of pure Python (often delegating low-level operations to libraries like NumPy and SciPy).
  \item Wrapping of existing (but previously unavailable for Python) libraries via Cython \cite{behnel10}.
  \item Direct implementation in Cython or C/C++.
\end{enumerate}

For the quantum transport solvers, a combination of approaches 1 and 2 is used.
There are currently two solvers, both based on solving the sparse linear system defined by Eqs.~(\ref{eq:tbham_inf},\ref{eq:scat_wave_func}): One utilizes the libraries UMFPACK \cite{umfpack1,umfpack2} or SuperLU \cite{superlu1,superlu2} that are wrapped by SciPy.
The other uses the more efficient MUMPS library \cite{mumps1,mumps2} that, however, is not included in SciPy and needs to be installed separately.
The MUMPS-based solver makes use of the nested dissection orderings provided by Metis \cite{metis} and Scotch \cite{scotch}.

The part of Kwant that calculates modes and self-energies of leads employs the same approach:
Pure Python code uses the services of the LAPACK library \cite{anderson99} that was specially wrapped with Cython as NumPy and SciPy do not provide all the necessary LAPACK functions.

Because highly optimized libraries perform the ``heavy lifting'' in these asymptotically most expensive parts of Kwant, the fraction of total running time spent in these libraries slowly approaches 100\% for all of Kwant as system size grows.
In practice more than half of all time is spent in such libraries for all but small systems which means that even if all of Kwant would have been implemented in highly optimized C or Fortran, the additional speed-up one could hope for would be less than a factor of two.
Given that this would be much more work, and that it would prevent one from profiting from the dynamic features of Python, we believe that such optimization would not be reasonable.

The routine \texttt{hamiltonian\_submatrix} that creates a representation of a (sub)matrix of the Hamiltonian of a Kwant system is an example of a component that has been optimized using Cython even though its execution time is only linear in terms of the number of sites (for sparse matrices).
The fact that this routine is called each time a system is solved for a given set of parameters makes it more performance-relevant than the construction and finalization a system that need to be performed only once for a given geometry.

The most low-level optimization of Kwant is embodied by the module \texttt{tinyarray} that has been implemented in pure C++ and is available as a stand-alone library for Python.
Kwant uses many instances of small vectors and matrices that cannot be collected into larger arrays, the most frequently used example being the tags and coordinates of lattice sites.
These small vectors and matrices could be represented by NumPy arrays or by Python tuples but both solutions are unsatisfactory:
NumPy arrays were not designed for this application and are hence too resource-hungry when used in great numbers.
Additionally, they cannot be used as keys for Python dictionaries.
Python tuples, on the other hand, do not provide mathematical operations and have a high memory-overhead as well, since each element of the tuple exists as an individual Python object.
The \texttt{tinyarray} library offers a solution by providing arrays that unlike those of NumPy are optimized for ``tiny'' sizes.
These arrays can be used as dictionary keys as they are \emph{hashable} and \emph{immutable}.

\section{Complete listings of the examples}
These example scripts can be downloaded from \url{http://downloads.kwant-project.org/examples/kwant-examples-njp.zip}.
\label{app:listings}
\subsection{Simplest builder usage}
\label{ex:simplest}
This example, explained in detail in Appendix~\ref{app:mappings}, constructs and plots (see Fig.~\ref{fig:simplest}) a simple system of three sites.
\begin{verbatim}
import matplotlib.pyplot
import kwant

lat = kwant.lattice.square()
sys = kwant.Builder()

# Add sites.
sys[lat(0, 0)] = 1.5
sys[lat(1, 0)] = 1.5
sys[lat(0, 1)] = 1.5

# Add hoppings.
sys[lat(0, 0), lat(1, 0)] = 2j
sys[lat(0, 1), lat(1, 0)] = 2j

kwant.plot(sys)
\end{verbatim}
\subsection{Circular quantum dot}
\label{ex:disk}
This example, featured in Appendix~\ref{app:iterables}, constructs and plots (see Fig.~\ref{fig:disk}) a disk-shaped quantum dot on a square lattice.
\begin{verbatim}
import matplotlib.pyplot
import kwant

lat = kwant.lattice.square()
sys = kwant.Builder()

def disk(pos):
    x, y = pos
    return x**2 + y**2 < 13**2

sys[lat.shape(disk, (0, 0))] = 0.5
sys[lat.neighbors(1)] = 1

kwant.plot(sys)
\end{verbatim}
\subsection{Irregularly shaped graphene quantum dot}
\label{ex:bean}
This example, featured in Appendix~\ref{app:iterables}, constructs and plots (see Fig.~\ref{fig:bean}) a ``bean''-shaped quantum dot on a honeycomb lattice.
\begin{verbatim}
import matplotlib.pyplot
import kwant

lat = kwant.lattice.honeycomb()
sys = kwant.Builder()

def bean(pos):
    x, y = pos
    rr = x**2 + y**2
    return rr**2 < 15 * y * rr + x**2 * y**2

sys[lat.shape(bean, (0, 1))] = 0.5
sys[lat.neighbors(1)] = 1

kwant.plot(sys)
\end{verbatim}
\subsection{System with leads}
\label{ex:leads}
This example, featured in Appendix~\ref{app:leads}, creates a ring-shaped finite system and a periodic infinite system.
The latter is attached twice as a lead to the ring: once outside and once inside.
Finally, the system is plotted (see Fig.~\ref{fig:leads}).
\begin{verbatim}
import matplotlib.pyplot
import kwant

lat = kwant.lattice.square()
sys = kwant.Builder()

def ring(pos):
    x, y = pos
    return 7**2 <= x**2 + y**2 < 13**2

sys[lat.shape(ring, (10, 0))] = 0
sys[lat.neighbors()] = 1

sym = kwant.TranslationalSymmetry((-2, 1))
lead = kwant.Builder(sym)
lead[(lat(x, 0) for x in range(-6, 6))] = 0
lead[lat.neighbors()] = 1
sys.attach_lead(lead)
sys.attach_lead(lead, lat(0, 0))

kwant.plot(sys)
\end{verbatim}
\subsection{Quantum Hall effect}
\label{ex:qhe}
This example creates a quantum point contact with two leads.
The system is subject to a perpendicular magnetic field and on-site disorder.
Quantum Hall effect plateaus are plotted (see Fig.~\ref{fig:qhe_plateaus}) that can be seen to break down with decreasing magnetic field strength.
A partially backscattered edge-state is plotted as well (see Fig.~\ref{fig:qhe_edgestate}).
Parts of this example are discussed in Appendix~\ref{app:values_ex_qhe} and Appendix~\ref{app:transport}.
\begin{verbatim}
import math
from cmath import exp
import numpy
from matplotlib import pyplot

import kwant
from kwant.digest import gauss

def hopping(sitei, sitej, phi, salt):
    xi, yi = sitei.pos
    xj, yj = sitej.pos
    return -exp(-0.5j * phi * (xi - xj) * (yi + yj))

def onsite(site, phi, salt):
    return 0.05 * gauss(repr(site), salt) + 4

def make_system(L=50):
    def central_region(pos):
        x, y = pos
        return -L < x < L and \
            abs(y) < L - 37.5 * math.exp(-x**2 / 12**2)

    lat = kwant.lattice.square()
    sys = kwant.Builder()

    sys[lat.shape(central_region, (0, 0))] = onsite
    sys[lat.neighbors()] = hopping

    sym = kwant.TranslationalSymmetry((-1, 0))
    lead = kwant.Builder(sym)
    lead[(lat(0, y) for y in range(-L + 1, L))] = 4
    lead[lat.neighbors()] = hopping

    sys.attach_lead(lead)
    sys.attach_lead(lead.reversed())

    return sys.finalized()

sys = make_system()
energy = 0.15

# Calculate and plot QHE conductance plateaus.
reciprocal_phis = numpy.linspace(4, 50, 200)
conductances = []
for phi in 1 / reciprocal_phis:
    smatrix = kwant.smatrix(sys, energy, args=[phi, ""])
    conductances.append(smatrix.transmission(1, 0))
pyplot.plot(reciprocal_phis, conductances)
pyplot.show()

# Calculate and plot a QHE edge state.
def density(sys, energy, args, lead_nr):
    wf = kwant.wave_function(sys, energy, args)
    return (abs(wf(lead_nr))**2).sum(axis=0)

d = density(sys, energy, [1/40.0, ""], 0)
kwant.plotter.map(sys, d)
\end{verbatim}
\subsection{Majorana Fermion}
\label{ex:majorana}
This example shows (see Fig.~\ref{fig:majorana}), as a function of magnetic field strength, the lowest eigenenergies of a system that supports Majorana fermions.
Parts of it are discussed in Appendix~\ref{app:values_ex_majorana} and Appendix~\ref{app:diag}.
\begin{verbatim}
from matplotlib import pyplot
import kwant
import numpy
import tinyarray
import scipy.sparse.linalg

# Python >= 3.3 provides SimpleNamespace in the
# standard library so we can simply import it:
# >>> from types import SimpleNamespace
# (Kwant does not yet support Python 3.)
class SimpleNamespace(object):
    """A simple container for parameters."""
    def __init__(self, **kwargs):
        self.__dict__.update(kwargs)

s_0 = numpy.identity(2)
s_z = numpy.array([[1, 0], [0, -1]])
s_x = numpy.array([[0, 1], [1, 0]])
s_y = numpy.array([[0, -1j], [1j, 0]])

tau_z = tinyarray.array(numpy.kron(s_z, s_0))
tau_x = tinyarray.array(numpy.kron(s_x, s_0))
sigma_z = tinyarray.array(numpy.kron(s_0, s_z))
tau_zsigma_x = tinyarray.array(numpy.kron(s_z, s_x))

def onsite(site, p):
    return tau_z * (p.mu - 2 * p.t) + \
        sigma_z * p.B + tau_x * p.Delta

def hopping(site0, site1, p):
    return tau_z * p.t + 1j * tau_zsigma_x * p.alpha

def make_system(l=70):
    sys = kwant.Builder()
    lat = kwant.lattice.chain()
    sys[(lat(x) for x in range(l))] = onsite
    sys[lat.neighbors()] = hopping
    return sys.finalized()

sys = make_system()

# Calculate and plot lowest eigenenergies in B-field.
B_values = numpy.linspace(0, 0.6, 80)
energies = []
params = SimpleNamespace(
    t=1, mu=-0.1, alpha=0.05, Delta=0.2)
for params.B in B_values:
    H = sys.hamiltonian_submatrix(
        args=[params], sparse=True)
    H = H.tocsc()
    eigs = scipy.sparse.linalg.eigsh(H, k=20, sigma=0)
    energies.append(numpy.sort(eigs[0]))
pyplot.plot(B_values, energies)
pyplot.show()
\end{verbatim}

\bibliography{kwant}

\end{document}